\documentclass[preprint,12pt]{elsarticle}

\usepackage{fancyhdr}
\fancypagestyle{pprintTitle}{%
\chead{}
\rhead{\vspace{-1in}NCU-HEP-k071 \\Feb. 2018\\ed.  Mar. 2019}
\lfoot{}
\cfoot{\thepage}
\rfoot{}

}

\usepackage{amsmath}
\usepackage{amssymb}
\usepackage{dsfont}
\usepackage{color}
\usepackage{graphicx,rotating,booktabs}

\def\bea{\begin{eqnarray}}
\def\eea{\end{eqnarray}}
\def\sea{\nonumber \\&&}
\def\lla{\left\langle}
\def\rra{\right\rangle}
\def\za{\alpha}
\def\zb{\beta}

\def\ssc{\scriptscriptstyle}
\def\lsim{\mathrel{\raise.3ex\hbox{$<$\kern-.75em\lower1ex\hbox{$\sim$}}} }
\def\gsim{\mathrel{\raise.3ex\hbox{$>$\kern-.75em\lower1ex\hbox{$\sim$}}} }

\begin{document}

\begin{frontmatter}

\title{The First Physics Picture of Contractions from a Fundamental Quantum Relativity Symmetry Including all Known Relativity Symmetries, Classical and Quantum}

\author{Otto C. W. Kong}
\ead{otto@phy.ncu.edu.tw}

\author{Jason Payne}

\address{ Department of Physics and  Center for High Energy and High Field Physics, 
%Center for Mathematics and Theoretical Physics, 
National Central University, Chung-li, Taiwan 32054  \\
}

%\cortext[cor1]{Corresponding author.}

\begin{abstract}
In this article, we utilize the insights gleaned from our recent 
formulation of space(-time), as well as dynamical picture of 
quantum mechanics and its classical approximation, from the 
relativity symmetry perspective in order to push further into 
the realm of the proposed fundamental relativity symmetry 
$SO(2,4)$. The latter has its origin arising from the perspectives 
of Planck scale deformations of relativity symmetries. We explicitly 
trace how the diverse actors in this story change through various 
contraction limits, paying careful attention to the relevant 
physical units, in order to place all known relativity theories 
-- quantum and classical -- within a single framework. More 
specifically, we explore both of the possible contractions of 
$SO(2,4)$ and its coset spaces in order to determine how best
 to recover the lower-level theories. These include both new 
models and all familiar theories, as well as quantum and 
classical dynamics with and without Einsteinian special 
relativity. Along the way, we also find connections with 
covariant quantum mechanics. The emphasis of this article 
rests on the ability of this language to not only encompass 
all known physical theories, but to also provide a path for 
extensions. It will serve as the basic background for more 
detailed formulations of the dynamical theories at each
 level, as well as the exact connections amongst them.
\end{abstract}

\end{frontmatter}
\thispagestyle{fancy}

\section{Introduction}
In the recent studies \cite{066,070}, we have given a successful formulation 
of quantum dynamics on a quantum space from a representation of what 
is essentially the $U(1)$ central extension of the Galilean symmetry 
$\widetilde{G}(3)$ or that of its $H_{\!\ssc R}(3)$ subgroup. The latter has 
one less generator, which is the time translation one. It is a semidirect 
product of the Heisenberg-Weyl group $H\!(3)$ with the $SO(3)$ rotations 
among the position observables $X_i$ and the momentum observables 
$P_i$. The dynamics part is a Heisenberg picture formulation on  the 
extension of the unitary representation of the group to one of the group 
$C^*$-algebra as the algebra of observables. Contraction of the
symmetry to trivialize the central extension gives the classical 
approximation. The contraction applied to the representations gives
Newtonian dynamics exactly \cite{070}. The formulation bases on the
the Hilbert space spanned by the canonical coherent states constructed
from the corresponding coset spaces of the relativity symmetry group.  
It is established that the projective Hilbert space can be justified to
be taken as a quantum model of the physical space, the classical limit
of which as given from the contraction can be seen as exactly the
Newtonian space model. Taking the latter challenging perspective 
or not, it is interesting to see if the approach can go further, like
incorporating Lorentz symmetry and beyond.

Since Einstein, physicists have learned to appreciate how intimately 
connected the notion of spacetime is with its relativity symmetry
 -- the symmetry of admissible reference frame /coordinate 
transformations. With the development of the mathematics of group 
theory and their representations, as well as their applications in physics 
(mostly in quantum mechanics), we have also learned to appreciate the 
perspective of taking the symmetry as the starting point for the 
construction of a theory.  In particular, both Newtonian space-time 
and Minkowski spacetime can be thought of as a (coset) representation 
space of the Galilean and Poincar\'e symmetry groups, respectively.  
The Newtonian model, as an approximation of the Einstein/Minkowski 
one, can be retrieved as a symmetry contraction limit \cite{IW,G,067}. 
The latter is just the mathematical way to implement, from the 
relativity symmetry perspective, the physical notion of taking the 
$c \to\infty$ limit, when all of the involved velocities are small 
compared to that of light ($c$). More interestingly, one can reason 
the other way around. The symmetry (algebra) of the Galilean boosts 
plus rotations is unstable against perturbations. When the vanishing  
structure constants in the commutator of two boost generators are 
made nonzero, one is forced to retrieve the (algebra of) the Lorentz 
symmetry as the unique (up to isomorphism) stabilized symmetry, 
with $\frac{1}{c}$ as the deformation parameter corresponding to 
an invariant speed $c$. Even if $c$ as never been measured to be 
finite, honest physicists can do nothing but give $\frac{1}{c}$ a lower 
bound. Whatever nonzero value for $\frac{1}{c}$ we admit, Lorentz 
symmetry provides one with a correct description of the relevant 
physics, while the Newtonian model can never be confirmed to be 
(exactly) correct -- merely correct up to some limitation in our 
measurements. Our proposed fundamental quantum relativity 
symmetry of $SO(2,4)$ \cite{030} comes from the idea of a fully 
stabilized symmetry, essentially incorporating all the known fundamental 
constants $G$, $\hbar$, and $c$ into the algebra structure, with and the 
Poincar\'e symmetry as (part of) a contraction limit. It is a formulation
of the deformed relativity approach within the Lie group/algebra
setting \cite{dsr,old}. For the sake of 
convenience, we note first the $SO(2,4)$ algebra is defined by
\bea \label{so} 
[J_{\ssc \!\mathcal R\mathcal S}, J_{\ssc\! \mathcal
M\mathcal N}] = -i ( \eta_{\ssc \mathcal S\mathcal M} J_{\ssc
\mathcal R\mathcal N} - \eta_{\ssc \mathcal R\mathcal M} J_{\ssc
\mathcal S\mathcal N} + \eta_{\ssc \mathcal R\mathcal N} J_{\ssc
\mathcal S\mathcal M} -\eta_{\ssc \mathcal S\mathcal N} J_{\ssc
\mathcal R\mathcal M}) \;,
\eea
where $\mathcal{R}$, $\mathcal{S}$, ${\mathcal M}$, and ${\mathcal N}$ 
range from 0 to 5, and we choose the metric convention 
$\eta_{\ssc \mathcal M\mathcal N} =( -1, 1, 1, 1, 1, -1)$. All $SO(m,n)$ 
symmetries involved in this article are (preserved) subgroups/subalgebras 
of $SO(2,4)$ for which we typically skip repeating the commutator 
structures among the $J$-generator sets. Our recent studies \cite{066,070}
put the relations between simple quantum mechanics and Newtonian
mechanics under the same perspective. The current article is about fully
connecting them all.

In Ref.\cite{060}, we looked into various contraction limits of the $SO(2,4)$ 
symmetry, as well as the corresponding contractions of the relevant coset 
space representations. The focus there was on symmetries that maintain 
a $SO(m,n)$ subgroup. $(m+n) \leq (2+4)$ is called the dimension of the 
relativity symmetry and the most interesting symmetries are obtained 
by contractions which take that dimension down one at a time.  The first 
step of the contraction is fixed as $SO(2,4) \to ISO(1,4)$ \cite{030}. 
Options for the further contractions from $ISO(1,4)$ were explored  
mathematically, while focusing on physically plausible pictures that result 
in, and hence go beyond, $G(1,3)$ -- a $(1+3)$D relativity symmetry of 
Galilean type. The present analysis, however, focuses on the contraction 
sequences passing through $H_{\!\ssc R}(1,3)$ instead. By this we mean 
a $(1+3)$D relativity symmetry of the Heisenberg-Weyl type. More 
specifically, it is a symmetry with generators $X_\mu$ and $P_\mu$, 
each transforming as components a four-vector under the $SO(1,3)$ 
subgroup generated by $J_{\mu\nu}$, together with a central charge 
generator giving the $X$-$P$ commutator a Heisenberg-type
 commutation relation. In total, the full $H_{\!\ssc R}(1,3)$ group 
therefore has 15 generators. Further contraction to some simple 
extension of an $H_{\!\ssc R}(3)$ group obviously works though 
it is not exactly trivial that the extended part causes no consistence 
problem with the known quantum physics. In fact, the whole
 $U(1)$ central extension $\widetilde{G}(3)$ of the Galilei group 
${G}(3)$ survives in the contracted symmetry we are after.

The $H_{\!\ssc R}(3)$ symmetry  is an invariant subgroup of $\widetilde{G}(3)$. 
 A sketch of the story is as follows: owing to the semidirect product structures 
\bea
\widetilde{G}(3)=H_{\!\ssc R}(3)\rtimes T_t =H(3) \rtimes (SO(3)\times T_t) \;,
\eea
where $T_t$ denotes the one parameter group of time translations, the standard Hilbert space representation of the Heisenberg-Weyl group $H(3)$ serves as a (spin zero) representation of $H_{\!\ssc R}(3)$, or $\widetilde{G}(3)$, in which the extra generators are simply represented by combinations of $\hat{X}_i$ and $\hat{P}_i$, i.e. the operators representing $X_i$ and $P_i$. The other relativity transformations act on $H(3)$ as outer automorphisms, as well as on its group algebra as inner automorphisms. The optimum framework for conceptual clarity is provided by formulating the Hilbert space as being spanned by the set of canonical coherent states $e^{i\theta} \!\left|p^i,x^i\rra$. Such states can 
themselves be identified with a coset space of $H_{\!\ssc R}(3)$ or $\widetilde{G}(3)$, and moreover in a way, as the group manifold of $H(3)$. The latter admits the 
coordinates $(p^i,x^i,\theta)$, with each group element given in the form $e^{i(p^i X_i - x^i P_i +\theta I)}$, where $I$ is the central charge. On the Hilbert space $\mathcal{K}$ of wavefunctions $\phi(p^i,x^i)=\lla p^i,x^i|\phi\rra$,
$\hat{X}_i$ and $\hat{P}_i$ are given by
\bea &&
x_i\star = x_i + i\partial_{p^i} \;,
\sea
p_i\star = p_i - i\partial_{x^i} \;,
\eea
where $\star$ is the Moyal star product (in the $\hbar=2$ units). The full algebra of observables is essentially the group $C^*$-algebra $C^*(H(3))$, represented as 
($L^{\!\ssc \infty}$) functions of the six basic operators listed above, $C(p_i\star, x_i\star) = C(p_i,x_i)\star$; or, equivalently, the multiplier algebra  
$\mathcal{M}'=\{\zb\in S':\zb\star\za \in L^2, \ \forall \za\in  L^2 \}$ represented as 
$I\!\!B(\mathcal{K})$. Observe that $\mathcal{K}$ actually sits inside the group algebra as the collection of partial isometries $\phi(p^i,x^i)\star$. Moreover, each real function $\za(p^i,x^i)$ gives rise to a Hermitian operator $\za(p^i,x^i)\star\equiv \za(p_i\star, x_i\star)$. Such an operator generates a one parameter group of unitary transformations on $\mathcal{K}$. The Heisenberg picture, envisaged as corresponding to a group of automorphisms on $C(p_i\!\star, x_i\!\star)$, can then be matched to the Schr\"odinger picture given by the latter description. The case in which $\za(p^i,x^i)\star$ is the energy operator yields time translation/evolution, and when the the energy operator is furthermore given by $\frac{p^ip_i}{2m}\star$, corresponding to a free particle, one obtains the Hamiltonian among the $\widetilde{G}(3)$ generators. 

The above description of quantum mechanics does not seem to offer any notion of the quantum configuration space, to say nothing of a quantum model of physical space. The Hilbert space $\mathcal{K}$, as a quantum phase space, is essentially the only variety of irreducible unitary representation of $H(3)$. Ref.\cite{066}, however, gives a clear justification for also interpreting $\mathcal{K}$ as a configuration space, from the perspective provided by the relevant coset space structures, as well as the relativity contraction limit trivializing the Heisenberg commutation relation yielding the classical/Newtonian approximation. The configuration space of a free particle is the only model of physical space one can have from any theory of particle dynamics; hence, from the quantum relativity perspective, the projective Hilbert space $\mathcal{P(K)}$, as an infinite-dimensional manifold, should be taken as the quantum model of physical space. Unlike the classical phase space, the quantum phase space as a representation of the its relativity symmetry is irreducible. The classical phase space is a sum of the configuration space and the momentum space, which is a splitting that cannot be made at the quantum level. As an irreducible representation of the quantum symmetry and the observable algebra, $\mathcal{K}$ becomes reducible upon the contraction. This can be seen as reducing the representation to the simple sum of the one-dimensional rays associated with the coherent state or the position eigenstate basis. Only such rays survive as pure (classical) states in the contraction limit. That is to say, the corresponding projective Hilbert spaces are exactly the phase space and (configuration) space cosets of the classical Galilean symmetry. 

This picture works well from the dynamical point of view, as well \cite{070}. Implementing the contraction on the observable algebra as the extension of the unitary representation of $H(3)$ to $C^*(H(3))$ results in the classical Poisson algebra. In other words, the deformation of $C(p^i,x^i)$ to $C(p^i\star,x^i\star)$, as in deformation quantization, is really a deformation of the corresponding relativity symmetry implemented on a representation of the relevant group $C^*$-algebra, which can be matched with a coset space representation and the corresponding unitary group representation (essentially on the Koopman-von Neumann Hilbert space of mixed states). The contraction is the `inverse' of the deformation, hence our formulation is one of \textit{dequantization}.

It is important to note, as illustrated in Ref.\cite{070}, that our picture of (quantum) relativity symmetries within the Lie group/algebra setting is a lot more powerful and generic than it may seem to be. The group $C^*$-algebra provides one with a noncommutative algebra to be taken as the observable algebra, on which the Lie group acts via automorphisms. It appears there is little reason to expect that the collection of all noncommutative algebras obtainable as the group $C^*$-algebra of some Lie group is not enough to describe the observable algebra of any fundamental physical theory we might have in mind. Moreover, to the extent that we would like to be able to retrieve some (quantum/noncommutative) spacetime picture out of it, we do expect a notion 
of relativity symmetry underlying everything. In physics, it sure looks as though we need little beyond the basic set of  phase space coordinate observables in order to describe all observables, though there may be a generalized notion of the latter beyond the classical position and momentum observables. This basic set is to be found among the generators of the relativity (Lie) algebra, while the full set is offered by the corresponding group $C^*$-algebra, interpreted as functions on this basic set. Thus, this basic set should be enough to fully illustrate the spacetime picture one is after. Specifically, the noncommutative geometry \cite{C} of the $C(p_i\!\star, x_i\!\star)$ algebra should be some manner of geometry equipped with the noncommutative coordinates $p_i\!\star$ and $x_i\!\star$. 

The success of the 3D quantum relativity picture naturally leads to the question of whether or not the analogous $(1+3)$D picture works as well. This question will be addressed below, together with the related question of how the 3D picture is to be retrieved from the $(1+3)$D picture as a relativity symmetry contraction limit. It is important to note that the ``dimension'' in both 3D and $(1+3)$D here is merely the dimension \textit{of the relativity symmetry}, which corresponds to the dimension of the corresponding space(time) in \textit{only the classical cases}. It can also likely be thought of as some variety of noncommutative dimension -- for instance, the 3D quantum relativity has three (noncommutative) $\hat{X}_i$ coordinate observables. Our first quantum space model in this setting is infinite-dimensional when thought of as a (commutative) manifold. Models for the higher levels, yet to be constructed, may even go completely beyond such real number geometric pictures.

The primary goal of the recent articles \cite{066,070} was to present a detailed picture of the feasibility of this whole scheme at the first level. We will illustrate here not only that a $(1+3)$D picture of what we have done in Refs. \cite{066,070} can be formulated, but also that the 
relevant coset space representations (of the $H_{\!\ssc R}(1,3)$ relativity symmetry) can be incorporated in sequences of representations starting from $SO(2,4)$, 
pass through $H_{\!\ssc R}(1,3)$, as well as including the extended symmetries of $H_{\!\ssc R}(3)$ or $\widetilde{G}(3)$, before eventually arriving at those relevant for Newtonian physics. We will, however, only briefly discuss the key notions relating to the full formulation of the associated dynamical theories, leaving such detailed investigations to be reported on in future publications. Moreover, we mostly leave such discussions until the end of the present article.

Let us elaborate a bit more on the basic framework of the program, especially in regards to one of the more challenging aspects that is crucial to the formulation 
and interpretation of the physical pictures at the various levels. The $SO(2,4)$ symmetry can be seen as arising from a stabilization of the algebra containing the Poincar\'e symmetry and the 3D Heisenberg algebra. Both from the perspective of our relativity symmetry stabilization, and that of requiring a consistent physical account for the relevant structures, we need to supplement these fourteen generators with an additional generator $X_{\ssc 0}$, promoting the Heisenberg structure to that of the $(1+3)$D version. While it looks like we can essentially use the Galilean boost $K_i$ generators as the position observable $X_i$ ($=\frac{1}{m} K_i$) at the 3D relativity symmetry level, we need the full $\widetilde{G}(3)$ symmetry (as the $U(1)$ central extension of the quantum relativity symmetry), which cannot be obtained from the Poincar\'e symmetry $ISO(1,3)$. In fact, the latter has one less generator. One can contract $ISO(1,3)$ to $G(3)$ \cite{067} or to $H_{\!\ssc R}(3) \equiv C(3)$ \cite{060}, but that is not enough. Similarly, in order to have a $(1+3)$D picture of both quantum and classical physics from this perspective, we need two different relativity symmetries connected by a contraction trivializing the Heisenberg commutation relation.  As the Poincar\'e symmetry does not even admit a nontrivial $U(1)$ central extension, this ten generator framework is certainly not enough. The $SO(2,4)$ symmetry is supposed to possess some manner of invariant length and invariant momentum, characterizing the noncommutativity among momentum and position observables, respectively, eventually. The $X_\mu$ generate what are called ``momentum boosts,'' which would be contracted to commuting momentum translations at the lower levels. We
have looked somewhat into the physics of such momentum boosts/translations in various settings \cite{030,060,023} (see also \cite{036,037}). The basic
feature that will likely be applicable in all cases is that $p^\mu$ has to be generally defined as the derivative of $x^\mu$ with respect to a new invariant parameter $\sigma$, {\em i.e.}  $p^\mu = \frac{\partial x^\mu}{\partial \sigma}$, and this is independent of the Newtonian idea of mass times velocity ($p^i=mv^i$).  In the case of an Einsteinian particle of mass $m$, we need to have $\sigma = \frac{\tau}{m}$, where $\tau$ is the particle's proper time, in order to retrieve the desired Einsteinian momentum expressions. Note that from the point of view of Hamiltonian mechanics, which is definitely the preferred setting for classical physics within the contraction formulation \cite{070}, even $p^i=mv^i$ is to be interpreted as an equation of motion, rather than a definition of momentum. 

From the phase space geometry point of view, at both classical and quantum levels, having momentum translations is as natural as having position translations, and they are, indeed, strongly suggested from the canonical coherent state picture. The crucial challenge that still needs to be fully appreciated is that such momentum translations would change the Einsteinian invariant particle mass. Moreover, we should be able to see this fact as a variety of reference frame transformation, though not necessarily one that is practically implementable. It is particularly interesting to note that the invariant parameter $\sigma$ has essentially been introduced in the theory of covariant quantum mechanics \cite{K,Fi}, which has a $(1+3)$D version of the Schr\"odinger equation. The latter theory is somewhat aligned with the basic spirit of the $(1+3)$D quantum relativity presented here. It is not, therefore, much of a surprise that that it is more or less the theory of quantum mechanics we obtained. Admitting such kinds of physical pictures as plausible theories -- only the limiting cases of which have been explored in the presently established cases -- seems to be very reasonable. Given that, we will illustrate below how our whole framework of quantum relativity looks quite promising, with step-by-step contractions, at least at the kinematical level. Furthermore, a first look at the dynamical setting will be discussed here. As we mentioned above, a fully dynamical analysis along the lines of Ref.\cite{070}, and the explicit contractions giving the lower levels as approximations, has to be left to future publications.

We begin, in the next section, with the picture of the $H_{\!\ssc R}(1,3)$ 
symmetry as a quantum relativity symmetry, and then briefly discuss the 
picture of covariant quantum mechanics that would result from the 
coherent state representations. This is based on the relevant cosets, 
in a fashion paralleling the case of the $H_{\!\ssc R}(3)$ symmetry. 
In Section 3, analysis of the potential contractions from $SO(2,4)$ 
illuminates which sequences of cosets give rise to those for the desired 
$H_{\!\ssc R}(1,3)$ symmetry. Particular attention is paid to the notion 
of physical dimensions, or the introduction of physical units, which can 
be seen as a consequence of the contractions. Section 4 deals with 
contractions applied to the Lorentz symmetry sitting within the 
$H_{\!\ssc R}(1,3)$ symmetry. The focus is on determining the 3D 
relativity symmetry pictures coming out of the quantum level first, 
and then taking them further down to the classical level. The quantum 
picture seems to be somewhat richer than that of the $H_{\!\ssc R}(3)$ 
or $\widetilde{G}(3)$ relativity symmetry -- some key physical issues 
related to this will be addressed in the last sections, after the analysis of 
the contractions of $H_{\!\ssc R}(1,3)$ to classical symmetries -- 
before further contracting the Lorentz symmetry -- in Section 5. 
Section 6 focuses particularly on the issue of physical dimensions for 
quantities and their relations at various levels of the relativity symmetries 
as traced through the contraction scheme. Interesting enough, it reveals 
a role of the Planck constant $\hbar$ different from the usual thinking. 
Finally, we provide some discussions and concluding remarks in Section 7. 
Note that we skip citations of the background references directly 
involved in analysis of the kind presented in Ref.\cite{070}, and 
leave it to interested readers to check the discussion and references 
contained therein.

\section{\boldmath $(1+3)$D Quantum Relativity Symmetry, Covariant Quantum Mechanics, and Classical Limits}
The quantum relativity symmetry perspective takes the Heisenberg commutation relation as a part of the relativity symmetry algebra. In order to have a similar formulation with the Lorentz symmetry of $SO(1,3)$ incorporated, the natural candidate to consider is that of the $H_{\!\ssc R}(1,3)$ symmetry. We will first highlight the particularly relevant coset space representations, in view of the analysis of Ref.\cite{066}. The nonzero commutators among the generators outside of the pure $SO(1,3)$ portion are taken to be
\begin{equation}
\begin{gathered}[]
[J_{\mu\nu}, X_\sigma]=
 -i(\eta_{\nu\sigma} X_\mu - \eta_{\mu\sigma} X_\nu) \;,
\qquad 
[J_{\mu\nu}, E_\sigma ] =
 -i(\eta_{\nu\sigma} E_\mu - \eta_{\mu\sigma} E_\nu) \;,
\\ 
[ X_\mu, E_\nu ] =2i \eta_{\mu\nu} I \;.
\end{gathered}\label{h13}
\end{equation}
Perhaps we should explain here our not-so-conventional notation. The last commutator is the $(1+3)$D Heisenberg commutation relation in which we use $E_\mu$, rather than $P_\mu$, and have an additional factor of $2$. The former is actually a natural feature of the relativity symmetry contraction picture, in which the introduction of
$P_\mu = \frac{1}{c} E_\mu$ with a nontrivial $c$ is really appropriate for taking the $c \to \infty$ contraction limit described below\footnote{\label{EvP}
%%%%%%%%%%\footnote{\label{fn}
One can introduce $P_\mu = \frac{1}{c} E_\mu$ and $I =  \frac{1}{c} F$, with $c$ being the speed of light, to write the algebra in terms of $J_{\mu\nu}$, $X_\mu$, 
$P_\mu$, and $I$ as generators. The latter form would be more familiar looking. Physics at that level would be better described in $c=1$ units anyway. Even a
simple contraction picture of the Poincar\'e to Galilean symmetry has the same feature. Interested readers can see Ref.\cite{067}, which gives a detailed pedagogical
description of the story.}.
%%%%%%%%%%%%%%%%%%%%%%
The factor of 2 arises from taking $\hbar=2$ units, which is the preferred choice of units for quantum mechanics. Note that the algebra would be the same under any independent changes of units for $X_\mu$ and $E_\mu$, as such variations can be absorbed into a redefined $I$. As $I$ commutes with everything else, it has to be
represented by a scalar multiple of the identity operator in any irreducible unitary representation. It is most convenient to choose a system of physical units such that $I$ in the above algebra can be taken as exactly the identity, and therefore $I$ is dimensionless. The physical dimension of $X_\mu$ and $E_\mu$ would then 
be reciprocals of one another, and an even better choice would be to take all of them as dimensionless.

The first coset of interest is obtained by factoring out the copy of $ISO(1,3)$ generated by the $J_{\mu\nu}$ and $X_\mu$ generators. The infinitesimal transformations of $H_{\!\ssc R}(1,3)$, or equivalently the action of the algebra element\footnote{
%%%%%%%%%%%%~\footnote{
 Strictly speaking, we should write the algebra elements with a factor of $-i$, which we leave out here and below.  The true generators of the real Lie algebra are really of the form $-iJ$, rather than simply $J$ itself. The conventional $-i$ is, of course, to have the `generators' $J$ represented by Hermitian operators in a 
unitary representation.
}
%%%%%%%%%%%%%
 $\frac{1}{2} \omega^{\mu\nu} J_{\mu\nu}
     + \bar{p}^\mu X_\mu -  \bar{\lambda}^{\mu} E_{\mu} +  \bar{\theta} I$ 
on the coset space coordinates $(\lambda^\mu, \theta)$, is given as follows: \vspace{.1in}
\\%%%%%%%%%%%%%%%%%%%%%%
{$\bullet$   $H_{\!\ssc R}(1,3)/ISO(1,3)$ :  ---}
\bea \label{hr13s}
\left(\begin{array}{c}
d\lambda^\mu \\ d\theta \\    0
\end{array}\right) =
\left(\begin{array}{ccc}
\omega^\mu_\nu & 0  &  \bar{\lambda}^\mu  \\
 2\bar{p}_\nu   &  0 &  \bar{\theta}\\
 {0}   & {0} &  0
\end{array}\right)
\left(\begin{array}{c}
{\lambda}^\nu \\ {\theta} \\    1
\end{array}\right)
=\left(\begin{array}{c}
\omega^\mu_\nu  \lambda^\nu  +\bar{\lambda}^\mu   \\ 
  2\bar{p}_\nu   \lambda^\nu  + \bar{\theta}\\   0
\end{array}\right) \;.
\eea\vspace{.015in}
\\%%%%%%%%%%%%%%%%%%%%%%
Another coset space of interest is $H_{\!\ssc R}(1,3)/SO(1,3)$, given by \vspace{.1in}
\\%%%%%%%%%%%%%%%%%%%%%%
$\bullet $ $H_{\!\ssc R}(1,3)/SO(1,3)$ : ---
\bea \label{hr13p}
\left(\begin{array}{c}
dp^\mu  \\ d\lambda^\mu\\ d\theta \\    0
\end{array}\right) =
\left(\begin{array}{cccc}
 \omega^\mu_\nu &  0& 0  &  \bar{p}^\mu\\
0  & \omega^\mu_\nu &  0 &  \bar{\lambda}^\mu\\
-\bar{\lambda}_\nu  & \bar{p}_\nu &  0 & \bar{\theta} \\
 {0}   & {0} &  0 & 0
\end{array}\right)
\left(\begin{array}{c}
p^\nu \\ \lambda^\nu  \\ \theta \\    1
\end{array}\right)
=\left(\begin{array}{c}
\omega^\mu_\nu   p^\nu + \bar{p}^\mu\\
\omega^\mu_\nu   \lambda^\nu + \bar{\lambda}^\mu \\ 
  \bar{p}_\nu   \lambda^\nu - \bar{\lambda}_\nu p^\nu + \bar{\theta} \\   0
\end{array}\right) \;.
\eea
These two cosets can be matched with their own associated coherent states, defined in terms of the unitary representation of operators on $H(3)$ given by
$e^{i\theta} \!\left|\lambda^\mu \rra
     = e^{i(\theta \hat{I} -\lambda^\mu \hat{E}_\mu)} \left|0 \rra$
and $e^{i\theta} \!\left|p^\mu,\lambda^\mu \rra
   = e^{i(\theta \hat{I} + p^\mu \hat{X}_\mu -\lambda^\mu \hat{E}_\mu)}\left|0,0 \rra$,
respectively. Each set of such states (without the phase factor) can be taken as a basis spanning a Hilbert space. Just as $\hat{E}_\mu$ effectively stands in for $\hat{P}_\mu$, $\lambda^\mu$ stands in for $x^\mu$ -- this means no more than expressing the same quantities in different physical units (see footnote \ref{EvP}). Taking $c=1$ units, one can simply identify each of these pairs. The two corresponding Hilbert space representations are, 
of course, equivalent. On each of the Hilbert spaces, as constructed above, a transformation (of the relativity group) takes a coherent state to another coherent state, exactly in accordance with the transformation between the corresponding coset space points. The $\left|\lambda^\mu \rra$ states are therefore eigenstates of $\hat{X}_\mu$ 
with eigenvalue $\lambda^\mu$. As such, we have a wavefunction representation of a generic state $\left|\phi\rra$ given by $\phi(\lambda^\mu) \equiv \lla \lambda^\mu | \phi\rra$, which are essentially the 
same as the wavefunctions found in covariant quantum mechanics \cite{K}. Wavefunctions on the canonical coherent state basis of $\left|p^\mu,\lambda^\mu \rra$, as given by $\phi(p^\mu,\lambda^\mu) \equiv \lla p^\mu,\lambda^\mu | \phi\rra$, can be acted on by $\hat{X}_\mu$ and $\hat{E}_\mu$ via the star product action
\bea &&
\hat{X}_\mu^{\!\ssc L} = \lambda_\mu\star = \lambda_\mu + i \partial_{p^\mu} \;,
\sea
\hat{E}_\mu^{\!\ssc L} = p_\mu\star = p_\mu - i \partial_{\lambda^\mu } \;,
\eea
with a generic operator from $C^*(H(1,3))$ considered as the operator function $\za(p_\mu\star,\lambda_\mu\star)=\za(p_\mu,\lambda_\mu)\star$. The latter is a great setting for the description of the dynamics in the Heisenberg picture and its contraction to the classical limit \cite{070}. The observables in the classical limit would be $\za(p_\mu^c,\lambda_\mu^c)$ functions of the classical phase space variables, the coset space picture of which will be obtained below. The Hilbert space becomes completely reducible as the sum of the one dimensional 
rays of the basis states. In other words, it effectively becomes the classical coset space in the contraction limit. In terms of the wavefunctions, only the coordinate delta functions $\delta(p_\mu^c,\lambda_\mu^c)$ survive as pure states \cite{066,070}.

The contraction to the classical limit is morally about decoupling $I$ -- removing it from any consideration of kinematics or dynamics. Fortunately, this is straightforward, as shown in Ref.\cite{066}. One simply introduces $X^c_\mu=\tfrac{1}{k} X_\mu$ and 
$E^c_\mu=\tfrac{1}{k} E_\mu$, and then takes the resulting commutation relations to the $k \to \infty$ limit. Apart from having  $X_\mu$ and $E_\mu$ replaced by $X^c_\mu$ and $E^c_\mu$, the algebra resulting from this further contraction only differs from the original by the now vanishing $[X^c_\mu, E^c_\nu]$. The resulting symmetry is that of $S(1,3)$ \cite{060}. If the $H_{\!\ssc R}(1,3)$ symmetry can be taken as the 
relativity symmetry for quantum physics on Minkowski spacetime, $S(1,3)$ would be the appropriate one for corresponding classical theory.  As an example, let us 
illustrate the contracted result of the first coset above as\vspace{.1in}
\\%%%%%%%%%%%%%%%%%
{$\bullet$   $S(1,3)/ISO(1,3)$ :  ---}
\bea \label{m4}
\left(\begin{array}{c}
d\lambda^\mu_c \\ d\theta \\    0
\end{array}\right) =
\left(\begin{array}{ccc}
\omega^\mu_\nu & 0  &  \bar{\lambda}^\mu_c  \\
 \frac{2}{k^2}\bar{p}_\nu   &  0 &  \bar{\theta}\\
 {0}   & {0} &  0
\end{array}\right)
\left(\begin{array}{c}
{\lambda}^\nu_c \\ {\theta} \\    1
\end{array}\right)
=\left(\begin{array}{c}
\omega^\mu_\nu  \lambda^\nu_c  +\bar{\lambda}^\mu_c   \\ 
  \bar{\theta}\\   0
\end{array}\right) \;.
\eea\vspace{.015in}

\noindent Apart from the completely decoupled $\theta$ coordinate, we have Minkowski space coordinated by the finite four-vector $\lambda^\mu_c =k \lambda^\mu$, which obey the relativity symmetry consisting of translations and Lorentz transformations. Note that for the other coset, $S(1,3)/SO(1,3)$, we also need $p^\mu_c = k p^\mu$: these are from the relations $p^\mu_c  X^c_\mu =p^\mu X_\mu$ and $\lambda^\mu_c  E^c_\mu =\lambda^\mu E_\mu$. 

As was mentioned in the introductory section, the physical picture with $H_{\!\ssc R}(1,3)$ and its classical limit is somewhat different from, or rather more general than, the limiting case as described by Einsteinian (special) relativity, and it is not our plan to give the full formulation and analysis of that case in this article. We will only give a brief sketch, drawing on references from the literature, to illustrate how such theories look very plausible as sensible generalizations of those based on Einsteinian relativity. In fact, some earlier efforts of our group developed a formulation of quantum and classical physics in a very similar setting \cite{036,037}, most parts of which are expected to still be feasible within the current framework. The setting presented in this earlier work is that of a $G(1,3)$ relativity -- obtained from an alternative path of contractions from $ISO(1,4)$ -- and its $U(1)$ central extension, {\em i.e.} a $\widetilde{G}(1,3)$ relativity. The quantum mechanics based on the latter can be seen as a geometric group quantization \cite{gq,dI} of the classical $G(1,3)$ theory. Then, based on the lessons of Ref.\cite{070}, we can infer that the basic quantum theory would mainly be a story of the  $H(1,3)$ invariant subgroup and its group $C^*$-algebra. To the extent that the rest of the generators are to be represented as some Hermitian function $\za(p_\mu\star,\lambda_\mu\star)$ generating inner automorphisms of the observable algebra, they have no significance in the formulation of the quantum dynamics. Their role is no different from any generic Hermitian observable, which also generates an automorphism, and the corresponding unitary transformation on the Hilbert space of $\phi(\lambda^\mu)$. In the case of the 3D version of $\widetilde{G}(3)$ and ${G}(3)$, the only special importance of the time translation generator is in the classical coset space representation that depicts Newtonian space-time, and the corresponding description of the Galilean boosts as space-time reference frame transformations. The analog of the latter here in our $(1+3)$D case is the parameter $\sigma$ mentioned in the introductory section and the generator giving its translations\footnote{
%%%%%%%%~\footnote{
Contractions of $ISO(m,n)$ to $G(m,n)$ or $H_{\!\ssc R}(m,n)$ (also commonly denoted by $C(m,n)$) differ only for one generator of the algebra, which plays the role of the time translation generator in $G(3)$ and the central charge needed for the Heisenberg commutation relation as in $H_{\!\ssc R}(3)$; for $G(1,3)$, it gives translations of an absolute (proper) time-like coordinate $\sigma$ in a five dimensional `spacetime' coset picture, besides the then relative coordinate time $t$, as in the $(1+3)$D Minkowski spacetime.
}.
%%%%%.
We surely can proceed happily without it and formulate the $\sigma$-translation only through an automorphism arising from the right generator. 

The naive quantum dynamical picture for states, as described by the wavefunctions $\phi(\lambda^\mu)$ (or more conventionally as $\phi(x^\mu)$ in the literature), is given by a proper time evolution of the form of the  Schr\"odinger equation with $-\frac{\hbar^2}{2}\hat{P}_\mu  \hat{P}^\mu$ as the Hamiltonian operator. Here, of course, we use $\hat{P}_\mu = -i\hbar \frac{\partial}{\partial {\lambda^\mu}}$. The lesson of the previous paragraph suggests that so long as we take the operator as the generic generator of a unitary flow, we get an associated `equation of motion' with the flow parameter as the `time'. The operator is to be taken as determining a version of free particle Hamiltonian $\tilde{G}(1,3)$ dynamics, the corresponding classical case of which retrieves exactly the particle dynamics as described by Einsteinian special relativity \cite{037}. The Klein-Gordon equation would be obtained as the `time' independent equation of motion where the particle rest mass (squared) has the role of an eigenvalue, rather than being an intrinsic characteristic of the particle. The `time' parameter in this case is $\sigma$, from which the Einstein proper time is given by dividing it by the mass. On the other hand, a formulation of Heisenberg picture dynamics as automorphism flow on 
$\za(p_\mu\star,\lambda_\mu\star)=\za(p_\mu,\lambda_\mu)\star$ would have the classical limit from the contraction given above given by Poisson/Hamilton dynamics, in agreement with that corresponding to $G(1,3)$. Therefore, we consider the $H_{\!\ssc R}(1,3)$ to $S(1,3)$ picture presented here to be convincingly acceptable at this stage.

It should be noted that the kind of Lorentz covariant formulation of classical and 
quantum mechanics has a long history. Readers can check, for example, the books 
by Fanchi \cite{Fi} and Trump and Schieve \cite{TS}, especially for the history and references.

\section{\boldmath The $H_{\!\ssc R}(1,3)$ Cosets from a Contraction of $SO(2,4)$}
Following Ref.\cite{030} closely would require the contraction to go through an intermediate $ISO(1,4)$; however, it is of interest to explore alternatives.
%%%%%%%%%%%%
\subsection{\boldmath $SO(2,4) \to ISO(1,4) \to H_{\!\ssc R}(1,3)$}
Let us first trace the contraction sequence explicitly, giving due attention to the important physical notion of the physical dimensions of quantities. From the Lie algebra of $SO(2,4)$, as given by Eq.(\ref{so}), we introduce the rescaled generators $E_{\ssc\!  A} = -\frac{1}{\lambda} J_{\ssc\!  A5}$,
$A$ from 0 to 4, and proceed to take them to the $\lambda \to \infty$ limit. This results in the following commutators
\bea
 && [J_{\ssc\!  AB},E_{\ssc\!  C}]
= -i (\eta_{\ssc  BC}E_{\ssc\!  A} - \eta_{\ssc  AC}E_{\ssc\!  B}) \;,
\nonumber   \\  &&
[E_{\ssc\!  A}, E_{\ssc\!  B} ]
= -\frac{i}{\lambda^2} J_{\ssc\!  AB} \rightarrow 0 \;.
\eea
A generic element of the algebra can be written as
\[
{\frac{1}{2} \omega^{\ssc \mathcal M\mathcal N} J_{\ssc\! \mathcal M\mathcal N}} 
= {\frac{1}{2} \omega^{\ssc  AB} J_{\ssc\!  AB}} - \lambda  \, \omega^{\ssc  A5} E_{\ssc\!  A} 
   \longrightarrow
  {\frac{1}{2}  \omega^{\ssc  AB} J_{\ssc\!  AB} } - \lambda^{\ssc  A} E_{\ssc\!  A}  \;,
\]
where the $\lambda^{\ssc  A} = \lambda \,\omega^{\ssc  A5}$ are taken to be finite in the $\lambda\to \infty$ limit. These $\lambda^A$ are the new 
parameters for the contracted algebra, which are matched to generators $E_{\ssc\!  A}$, and they have the physical dimensions $[\lambda]^{-1}$ (while $\lambda^{\ssc  A}$ has dimension $[\lambda]$). The resulting symmetry is 
$ISO(1,4)$. Note that the natural choice of physical units in the $SO(2,4)$ case is no unit at all, {\em i.e.} all $J_{\ssc\! \mathcal M\mathcal N}$ and $\omega^{\ssc \mathcal M\mathcal N}$ are dimensionless. The physical 
meaning of the $\lambda \to \infty$ contraction as an approximation \cite{G,060}, however, says that observables corresponding to the $E_{\ssc\!  A}$ generators appear to be different kinds of physical quantities than the $J_{\ssc\!  AB}$. Reflecting on the physical meaning of this statement about the $ISO(1,4)$ 
symmetry, physicists would introduce a physical unit for $E_{\ssc\!  A}$ (different from that of $J_{\ssc\!  AB}$) -- the actual finite numerical value of $\lambda$ with respect to this unit would be taken as a fundamental 
constant.  That is, for example, the nature of the speed of light $c$ \cite{067}.

For the second contraction, we want to separate $E_{\ssc  4}$ from the remaining $E_\mu$, and rescale either one along with the  $J_{\mu\ssc 4}$. Again, the idea is to reduce the dimension of the relativity symmetry by one, specifically from $(1+4)$ to $(1+3)$.  Summarizing the results in Ref.\cite{060}: taking $E_\mu$ through this contraction yields the aforementioned $G(1,3)$, while rescaling $E_{\ssc  4}$ instead results in the $C(1,3)$ symmetry; performing neither or both gives $S(1,3)$, which itself can be obtained by a further contraction from $G(1,3)$ or $C(1,3)$. The $C(1,3)$ notation comes from $C(3)$ -- the so-called Carroll symmetry that has been around in relativity symmetry discussions for 
some time (see for example Ref.\cite{BL}), but apparently without realizing it has anything to do with quantum mechanics. Mathematically, the Carroll symmetry is really just what we have denoted by $H_{\!\ssc R}$, and as such we will drop the $C(m,n)$ notation in favor of $H_{\!\ssc R}(m,n)$ for the remainder of this paper.  An explicit illustration of the contraction $ISO(1,4) \to H_{\!\ssc R}(1,3)$ can  be given as the $p \to \infty$ limit applied to $X_\mu= \frac{1}{p} J_{\mu\ssc 4}$ and $F= \frac{1}{p} E_{\ssc 4}$; we then obtain (skipping the $[J,J]$ part)
\bea
&& [J_{\mu\nu}, X_\sigma ] =
 -i(\eta_{\nu\sigma} X_\mu - \eta_{\mu\sigma} X_\nu) \;,
\qquad
%\nonumber \\ &&
[ X_\mu, X_\nu ] = \frac{i}{p^2}J_{\mu\nu} \to 0 \;,
\nonumber \\ &&
[J_{\mu\nu}, E_\sigma ] =
 -i(\eta_{\nu\sigma} E_\mu - \eta_{\mu\sigma} E_\nu) \;,
\quad\quad
%\nonumber \\ &&
[ E_\mu, E_\nu ] =  0 \;,
\quad\quad 
[ E_\mu, F] =  0 \;,
\nonumber \\ &&
[ X_\mu, E_\nu ] = i\eta_{\mu\nu} F \;,
\quad\quad
[ X_\mu, F ] = -\frac{i}{p^2}E_\mu \to 0 \;,
\qquad
%\nonumber \\ &&
[J_{\mu\nu}, F]=0 \;, \label{iso/hr}
\eea
which, upon identifying $F$ with $2I$, is exactly the algebra of Eq.(\ref{h13}).
%******* {\bf keep $[X,F]$ violate JI  (X,X,P)}\\

Once again, we want to trace the (relative) physical dimensions of the quantities represented by the generators through the contraction. One can see that $E_\mu$ possesses the dimensions $[\lambda]^{-1}$, while $X^\mu$ has $[p]^{-1}$, and $F$ carries the dimensions of $[\lambda]^{-1}[p]^{-1}$. Obviously, the dimension of $F$ is that of $\hbar$. Generic elements of the algebras are related as follows:
\bea
%{\frac{1}{2}J_{\ssc\! \mathcal M\mathcal N}\omega^{\ssc \mathcal M\mathcal N}} 
%&& =
%{\frac{1}{2}J_{\ssc\!  AB}\omega^{\ssc  AB}} - E_{\ssc\!  A} \,\lambda  \, \omega^{\ssc  A5}
%\nonumber \\ 
%\longrightarrow &&
{\frac{1}{2} \omega^{\ssc  AB} J_{\ssc\!  AB} } - \lambda^{\ssc  A} E_{\ssc\!  A} 
&& \hspace{-.25in}=  { \frac{1}{2}  \omega^{\mu\nu} J_{\mu\nu}}+ p \,\omega^{\mu\ssc  4} X_\mu 
   - \lambda^{\mu} E_{\mu}   - p \lambda^{\ssc 4} F 
\sea
\hspace{-.25in}\longrightarrow 
  {\frac{1}{2} \omega^{\mu\nu} J_{\mu\nu}} + p^\mu X_\mu - \lambda^{\mu} E_{\mu}  + f F \;.
%\nonumber \\ && %\hspace*{2in} 
%\left( = {\frac{1}{2}J_{\mu\nu}\omega^{\mu\nu}} + X_\mu \, \bar{\kappa}^\mu + P_{\mu}  \,x^{\mu} + I s \right) .
\eea
Note that  $p^\mu = p \, \omega^{\mu\ssc  4}$  and $f =-p\,\lambda^{\ssc 4}$  
%(with $x^\mu = c \lambda^\mu$ and $s= c f$) 
are the finite parameters at the corresponding contraction limits,
 with physical dimensions of $[p]$ and $[\lambda][p]$, respectively.

The sequence of symmetry contractions can moreover be implemented on the coset spaces $SO(2,4)/SO(1,4)$ and $SO(2,4)/SO(1,3)$, which are in turn contracted as
\[
SO(2,4)/SO(1,4) \to ISO(1,4)/SO(1,4) \to H_{\!\ssc R}(1,3)/ISO(1,3) \;,
\]
and
\[
SO(2,4)/SO(1,3) \to ISO(1,4)/SO(1,3) \to H_{\!\ssc R}(1,3)/SO(1,3) \;.
\]
%%%%%%%%%%%%%%%%%%%%%%%%%%%%%
The coset space $SO(2,4)/SO(1,4)$ can be considered as the hyperbolic space, sitting inside of the $(2+4)$D pseudo-Euclidean space with coordinates $ Z^{\!\ssc \mathcal M}$, defined by the condition
\[
\eta_{\ssc \mathcal M\mathcal N} Z^{\!\ssc \mathcal M} Z^{\!\ssc \mathcal N} =-1 \;.
\]
The infinitesimal action of $SO(2,4)$ on these coordinates is simply given by 
\bea
d Z^{\!\ssc \mathcal M} = \omega^{\ssc \mathcal M}_{\ssc \mathcal N} Z^{\!\ssc \mathcal N} \;.
\eea

In order to see how the contraction can be implemented more explicitly, let us first rewrite the above as \vspace{.1in}
\\%%%%%%%%%%%%%%%%%%%%%%
 { $\bullet$ $SO(2,4)/SO(1,4)$: ---}  
\bea\label{24c}
\left(\begin{array}{c}
dZ^{\!\ssc A} \\ dZ^{\ssc 5}
\end{array}\right)
=\left(\begin{array}{cc}
\omega^{\ssc A}_{\ssc B} &  \omega^{\ssc A}_{\ssc 5} \\
 \omega^{\ssc 5}_{\ssc A}  & 0
\end{array}\right)
\left(\begin{array}{c}
 Z^{\!\ssc B} \\ Z^{\ssc 5}
\end{array}\right)
=\left(\begin{array}{cc}
\omega^{\ssc A}_{\ssc B} &  \frac{-1}{\lambda} \bar\lambda^{\ssc A}\\
 \frac{-1}{\lambda} \eta_{\ssc BA }\bar\lambda^{\ssc A} & 0
\end{array}\right)
\left(\begin{array}{c}
 Z^{\!\ssc B} \\ Z^{\ssc 5}
\end{array}\right) ,
\eea
where following the above, we use  
$\bar\lambda^{\ssc A}$ ($=\lambda\omega^{\ssc A5}$) in place of $\lambda^{\ssc A}$ in the transformation, mostly to distinguish it from the latter in the coset space coordinates below. One merely needs to use the
new coordinates $\lambda^{\ssc A} =\lambda Z^{\!\ssc A}$, focused on the region around the point $Z^{\ssc 5}=-1$, and then take the picture to the $\lambda \to \infty$ limit.  $\bar{\lambda}^{\ssc A}$ becomes what is essentially a translation of $\lambda^{\!\ssc A}$. The following would be the result of this.
\vspace{.1in}
\\%%%%%%%%%%%%%%%%%%%%%%
 {$\bullet$  $ISO(1,4)/SO(1,4)$: ---}  
\bea\label{14s} &&\!\!\!\!\!\!\!\!
\left(\begin{array}{c}
d\lambda^{\ssc A} \\   d(1)
\end{array}\right) \!\!
= \!\! \left(\begin{array}{cc}
\omega^{\ssc A}_{\ssc B} &  \bar{\lambda}^{\ssc A} \\
 \frac{1}{\lambda^2} \bar{\lambda}_{\ssc B}  & 0
\end{array}\right) \!\!
\left(\begin{array}{c}
 \lambda^{\ssc B} \\ 1
\end{array}\right)
= \left(\begin{array}{c}
\omega^{\ssc A}_{\ssc B} \lambda^{\ssc B}  +  \bar{\lambda}^{\ssc A}\\ 
  \frac{1}{\lambda^2} \bar{\lambda}_{\ssc B} \lambda^{\ssc B} 
\end{array}\right) 
\to \left(\begin{array}{c}
\omega^{\ssc A}_{\ssc B} \lambda^{\ssc B}  +  \bar{\lambda}^{\ssc A}\\  0
\end{array}\right).
\sea
\eea
This is exactly the structure we have for the $ISO(1,4)/SO(1,4)$ coset, akin to a five dimensional Minkowski space. Contraction pictures of this kind are quite standard, and are well-described with a geometric language \cite{G}. 

Another account based on more physical reasoning can be given as follows: for large values of $|Z^{\ssc 5}|$, probably with all $|Z^{\!\ssc A}|$  large as well, we can take approximation of the $SO(2,4)$ coset space as satisfying $\eta_{\ssc \mathcal M\mathcal N} Z^{\!\ssc \mathcal M} Z^{\!\ssc \mathcal N} =0$; hence, it is no longer curved. This is equivalent to  $\eta_{\!\ssc AB} Z^{\!\ssc A} Z^{\!\ssc B} =|Z^{\ssc 5}|^2$, the latter of which can be taken as a free (positive) parameter, and we can forget about $Z^{\ssc 5}$ as a coordinate. One would then describe all of the large numerical values of $Z^{\!\ssc A}$ by some $\lambda^{\ssc A}$, with a convenient choice of physical units, which reduces the size of their numerical values. This is in much the same vein as how the Planck length provides a natural length scale for nature compared to which the usual scale of laboratory physics is essentially infinite. Planck length is expected to be a notion of minimal length, close to the scale of which we expect very nontrivial structure of spacetime. In the `normal' setting of laboratory physics, we can neglect this and see that the notion of metric distance has no lower bound. This is precisely the spirit behind introducing a relativity deformation to $SO(2,4)$, and the above contraction is the `inverse' of this deformation.

Going from $ISO(1,4)/SO(1,4)$  to $H_{\!\ssc R}(1,3)/ISO(1,3)$ is straightforward, we have:\vspace{.1in}
\\%%%%%%%%%%%%%%%%%%%%%%
 {$\bullet$  $ISO(1,4)/SO(1,4) \to H_{\!\ssc R}(1,3)/ISO(1,3)$: ---}  
\bea\label{c13s}
\left(\begin{array}{c}
d\lambda^\mu \\  df \\    0
\end{array}\right) =
\left(\begin{array}{ccc}
\omega^\mu_\nu & -\frac{1}{p^2} \bar{p}^\mu  &  \bar{\lambda}^\mu  \\
       \bar{p}_\nu   &  0 &   \bar{f}\\
 {0}   & {0} &  0
\end{array}\right)
\left(\begin{array}{c}
  \lambda^\nu \\ {f} \\    1
\end{array}\right)
=\left(\begin{array}{c}
\omega^\mu_\nu   \lambda^\nu  +   \bar{\lambda}^\mu   \\ 
 \bar{p}_\nu   \lambda^\nu  +    \bar{f}\\   0
\end{array}\right) \;,
\eea
in which we have used $f$ and $\bar{p}^\mu$ instead of $\lambda^{\ssc 4}$ and $\omega^{\mu\ssc 4}$. The result is to be read at the $p\to\infty$ limit, though we show the terms vanishing in the limit in the calculation so that readers can easily trace how the result is obtained (a presentational feature that we will maintain below). The coset is, of course, the same as the somewhat 
differently written Eq.(\ref{hr13s}).

Similar considerations to those given above lead us to write: \vspace{.1in}
\\%%%%%%%%%%%%%%
 {$\bullet$  $SO(2,4)/SO(1,3) \to ISO(1,4)/SO(1,3) $ ---}  
\bea %\label{14p}
\left(\begin{array}{c}
d\lambda^{\ssc A} \\  d(1)\\ dY^{\!\mu} \\    dY^{\!\ssc  4}
\end{array}\right) =
\left(\begin{array}{cccc}
\omega^{\ssc A}_{\ssc B}  &    \bar{\lambda}^{\ssc A}  &  0  &  0 \\
\frac{1}{\lambda^2} \bar{\lambda}_{\ssc B} &  0  &  0 & 0 \\
0  & 0  & \omega^\mu_\nu  &  {\omega}^{\mu}_{\ssc  4}\\
 %\frac{1}{2}\bar{x}_\nu  & - \frac{1}{2}\bar{p}_\nu &  0 & \bar{f} \\
  0  & 0  &      {\omega}_\nu^{\ssc  4} &  0 
\end{array}\right)
\left(\begin{array}{c} 
\lambda^{\ssc B}  \\ 1 \\ Y^{\!\nu}  \\   Y^{\!\ssc  4}
\end{array}\right)
=\left(\begin{array}{c}
\omega^{\ssc A}_{\ssc B}   \lambda^{\ssc B}  + \bar{\lambda}^{\ssc A}  \\ 0 \\
\omega^\mu_\nu  Y^\nu +  {\omega}^{\mu}_{\ssc  4} Y^{\!\ssc  4}
 \\ {\omega}_\nu^{\ssc  4} Y^{\!\ssc  4}
 %\frac{1}{\lambda^2} \bar{\lambda}_{\ssc B} \lambda^{\ssc B} 
\end{array}\right) ,
\eea
which is simply the sum of two parts, namely $ISO(1,4)/SO(1,4)$, as described above in Eq.(\ref{14s}), and $SO(1,4)/SO(1,3)$, described along the lines of Eq.(\ref{24c}) in terms of the $Y^{\!\ssc A}$ coordinates.
It is then straightforward to take a further contraction, essentially along the lines of taking $SO(1,4)/SO(1,3)$ to $ISO(1,3)/SO(1,3)$, and similar to the discussion above (here with $p^\mu =p Y^\mu$ and $Y^{\!\ssc  4} \sim 1$). The resulting coset space is given by \vspace{.1in}
\\%%%%%%%%%%%%%%
{$\bullet$  $ISO(1,4)/SO(1,3) \to H_{\!\ssc R}(1,3)/SO(1,3)$ ---}
\bea \label{c13p}
\left(\begin{array}{c}
dp^\mu  \\ d\lambda^\mu\\ df \\    d(1)
\end{array}\right) =
\left(\begin{array}{cccc}
 \omega^\mu_\nu &  0& 0  &  \bar{p}^\mu\\
0  & \omega^\mu_\nu & 0 &  \bar{\lambda}^\mu\\
 0 &  \bar{p}_\nu &  0 & \bar{f} \\
 \frac{1}{p^2}  \bar{p}_\nu   & {0} &  0 & 0
\end{array}\right)
\left(\begin{array}{c}
p^\nu \\ \lambda^\nu  \\ f \\    1
\end{array}\right)
=\left(\begin{array}{c}
\omega^\mu_\nu   p^\nu + \bar{p}^\mu\\
\omega^\mu_\nu   \lambda^\nu  + \bar{\lambda}^\mu \\ 
    \bar{p}_\nu   \lambda^\nu  + \bar{f} \\   0
\end{array}\right) \;,
\eea
which can be seen as giving essentially the same physical picture as Eq.(\ref{hr13p}). We leave details of the various issues one might be concerned with to be discussed below.

\subsection{\boldmath $SO(2,4) \to ISO(2,3) \to H_{\!\ssc R}(1,3)$ \textnormal{\textbf{and}} $SO(2,4)  \to H_{\!\ssc R}(1,3)$ \textnormal{\textbf{Directly}}}
$ISO(2,3) $ is an obvious alternative symmetry between $SO(2,4)$ and $H_{\!\ssc R}(1,3)$, the contraction sequence of which should not be expected to be much different from the one given above passing through $ISO(1,4)$. We sketch it briefly here to address any differences that may be worth some attention.

Basically, one has to take the two contractions in reverse order, namely taking
$X_{\mu}= \frac{1}{p} J_{\mu\ssc 4}$ and $X_{\ssc 5}= -\frac{1}{p} J_{\ssc\! 45}$, for the first step, and then $E_\mu=-\frac{1}{\lambda} J_{\mu\ssc  5}$ and
$F=\frac{1}{\lambda}X_{\ssc 5}$ for the second. The cosets of $ISO(2,3)/SO(2,3)$ and $ISO(2,3)/SO(1,3)$, obtainable from the contraction of  $SO(2,4)/SO(2,3)$ and 
$SO(2,4)/SO(1,3)$, respectively, have essentially the same basic structure as $ISO(1,4)/SO(1,4)$ and $ISO(1,4)/SO(1,3)$. Note, however, that $ISO(2,3)/SO(2,3)$ is a pseudo-Euclidean space with signature $\{-1,1,1,1,-1\}$, and the five vectors have as coordinates $p^\mu$ and $p^{\ssc 5}= p\, \omega^{\ssc 5\!4}$, instead of $\lambda^{\ssc A}$.  For $H_{\!\ssc R}(1,3)/ISO(1,3)$, obtained from the latter, we have\vspace{.1in}
\\%%%%%%%%%%%%%%%%%%%%%%
{$\bullet$    $H_{\!\ssc R}(1,3)/ISO(1,3)$ from $ISO(2,3)/SO(2,3)$ : ---}  
\bea\label{c13sa}
\left(\begin{array}{c}
dp^\mu \\  df \\    0
\end{array}\right) =
\left(\begin{array}{ccc}
\omega^\mu_\nu & \frac{1}{\lambda^2} \bar{\lambda}^\mu  &  \bar{p}^\mu  \\
      -  \bar{\lambda}_\nu   &  0 &   \bar{f}\\
 {0}   & {0} &  0
\end{array}\right)
\left(\begin{array}{c}
  p^\nu \\ {f} \\    1
\end{array}\right)
=\left(\begin{array}{c}
\omega^\mu_\nu   p^\nu  +   \bar{p}^\mu   \\ 
 -  \bar{\lambda}_\nu   p^\nu  +    \bar{f}\\   0
\end{array}\right) \;.
\eea
One should observe that we have the same expression 
$df= -  \bar{\lambda}_\nu   p^\nu  +    \bar{f}$ for the
$H_{\!\ssc R}(1,3)/SO(1,3)$ coset coming from $ISO(2,3)/SO(1,3)$, which is however different from those given in the above presentation of the $H_{\!\ssc R}(1,3)/SO(1,3)$ coset. The coset itself should obviously be the same as the one obtained from $ISO(1,4)/SO(1,4)$ as we have the same $H_{\!\ssc R}(1,3)$ group and $SO(1,3)$ subgroup. The difference in the explicit transformation of $f$, which corresponds to the phase of a state in the associated quantum mechanics being described, can in fact be appreciated via a $U(1)$ central extension analysis \cite{dI}. The group generated by the Heisenberg algebra can be written with formally different group products that are related by some cocycle. This issue also explains the different forms given in the coset presentations of the previous section compared with those in this section. The $H_{\!\ssc R}(1,3)/ISO(1,3)$ coset is, however, not really the same as the other ones above, as the $ISO(1,3)$ subgroups we are concerned with here are really different from the ones with generators originating as $J_{\mu\nu}$ and $J_{\mu\ssc 5}$, and $J_{\mu\nu}$ and $J_{\mu\ssc 4}$ in the other cases. We will not attempt to formulate any dynamical models for physics above the $H_{\!\ssc R}(1,3)$ level at this point, and will leave the issues concerning dynamics for future investigation. At the $H_{\!\ssc R}(1,3)$ level, it is obvious that we have something like the configuration space coset in one case, and something like the momentum space coset in the other, which are also precisely what we get upon further contractions to decouple the quantum central charge $F$.
% f=-\lambda^4 p = p^5 \lambda=-{p\lambda} \omega^{45}
%\bea
%{\frac{1}{2}J_{\ssc\! \mathcal M\mathcal N}\omega^{\ssc \mathcal M\mathcal N}} 
%&& =
%{\frac{1}{2}J_{\ssc\!  AB}\omega^{\ssc  AB}} - E_{\ssc\!  A} \,\lambda  \, \omega^{\ssc  A5}
%\nonumber \\ 
%\longrightarrow &&
%{\frac{1}{2} \omega^{\ssc  AB} J_{\ssc\!  AB} } - \lambda^{\ssc  A} E_{\ssc\!  A} 
 %=  { \frac{1}{2}  \omega^{\mu\nu} J_{\mu\nu}}+ p \,\omega^{\mu\ssc  4} X_\mu 
 %  - \lambda^{\mu} E_{\mu}   - p \lambda^{\ssc 4} F 
%\nonumber \\ 
%\longrightarrow &&
 % {\frac{1}{2} \omega^{\mu\nu} J_{\mu\nu}} + p^\mu X_\mu - \lambda^{\mu} E_{\mu}  - f F \;.
%\sea
 %=  { \frac{1}{2}  \omega^{\mu\nu} J_{\mu\nu}} +  p \,\omega^{\mu\ssc  4} X_\mu 
%+  p \,\omega^{\ssc  54} X_5  +  \omega^{\mu\ssc  5}  J_{\mu\ssc 5}
%\sea \longrightarrow 
%+  p^5  X_5 +  \omega^{\mu\ssc  5}  J_{\mu\ssc 5}
%= { \frac{1}{2}  \omega^{\mu\nu} J_{\mu\nu}} +  p^\mu X_\mu 
%\eea

One other alternative that is actually more interesting is to contract $SO(2,4)$ to $H_{\!\ssc R}(1,3)$ directly, which can be achieved by taking $E_{\mu} = -\frac{1}{\lambda} J_{\mu\ssc 5}$, $X_\mu=\frac{1}{p} J_{\mu\ssc 4}$, and $F=-\frac{1}{p\lambda} J_{\ssc 45}$ simultaneously to the $\lambda,p \to \infty$ limit. This is more naturally done by simply identifying $\lambda$ and $p$.  We keep them separate here mostly for easy comparison with the two-step contraction pictures. One particularly noteworthy point is that for the phase space coset of $H_{\!\ssc R}(1,3)/SO(1,3)$ the two alternative sequences of contractions described above give different 
expressions for $df$: $f$ from the contraction of $\lambda^{\ssc 4}$ yields a $\bar{p}_\nu \lambda^\nu$ contribution while $F$ from $p^{\ssc 5}$ results in $-  \bar{\lambda}_\nu   p^\nu$, both obviously a consequence of the nontrivial $X_\mu$-$E_\nu$ commutation relation. It is easy to appreciate the fact that taking the single step contraction from $SO(2,4)$ should not show any preference for one of the two expressions over the other; hence a more symmetric form of Eq.(\ref{hr13p}) is to be expected, {\em i.e.}, with $df=\frac{1}{2}(\bar{p}_\nu   \lambda^\nu-  \bar{\lambda}_\nu   p^\nu) +\bar{f}$.

To formulate the picture of the passage of $SO(2,4)/SO(1,3)$ to $H_{\!\ssc R}(1,3)/SO(1,3)$ along the contraction, one can use a description of the first coset space by a set of eleven coordinates: $Z^{\mu}$, $Z^{\ssc 5}$, $Y^{\!\ssc A}$ and $W$, with $(Z^{\mu},\frac{-W}{Y^{\ssc  5}}, Z^{\ssc 5})$ and $(Y^{\!\ssc A},\frac{-W}{Z^{\ssc 4}})$ 
transforming as six-vectors under $SO(2,4)$, {\em i.e.} 
$W= -Z^{\ssc 4} Y^{\ssc 5}$. We have%\vspace{.1in}
\bea %\label{14p}
\left(\begin{array}{c}
dZ^{\mu} \\ dZ^{\ssc 5} \\  dW \\    dY^{\!\ssc  4} \\ dY^{\!\mu}
\end{array}\right) =
\left(\begin{array}{ccccc}
\omega^\mu_\nu  & \omega^{\mu}_{\ssc 5}
 &    \frac{-1}{Y^{\ssc 5}}\omega^{\mu}_{\ssc 4} &  0  &  0 \\
 \omega_{\nu}^{\ssc 5} &  0  & \frac{-1}{Y^{\ssc 5}}\omega^{\ssc  5}_{\ssc 4} 
&  0 & 0 \\
-Y^{\ssc 5} \, \omega_{\nu}^{\ssc 4} & -Y^{\ssc 5} \,  \omega^{\ssc  4}_{\ssc 5}   & 0
&  -Z^{\ssc 4}\, \omega^{\ssc  5}_{\ssc 4} & -Z^{\ssc 4}\, \omega_{\nu}^{\ssc 5}  \\
0  & 0  &   \frac{-1}{Z^{\ssc 4}}\omega^{\ssc  4}_{\ssc 5} & 0 &   {\omega}_\nu^{\ssc  4}   \\
 %\frac{1}{2}\bar{x}_\nu  & - \frac{1}{2}\bar{p}_\nu &  0 & \bar{f} \\
  0  & 0  &   \frac{-1}{Z^{\ssc 4}}{\omega}^{\mu}_{\ssc  5} &  {\omega}^{\mu}_{\ssc  4}  
  & \omega^\mu_\nu 
\end{array}\right)
\left(\begin{array}{c} 
Z^{\nu}  \\   Z^{\ssc  5}\\ W   \\   Y^{\!\ssc  4} \\ Y^{\!\nu}
\end{array}\right) .
\eea
This coset description is really just putting together the $SO(2,4)/SO(1,4)$ coset picture of $Z^{\!\mathcal M}$ with $\eta_{\ssc \mathcal M\mathcal N} Z^{\!\ssc \mathcal M} Z^{\!\ssc \mathcal N} =-1$, and the $SO(2,4)/SO(2,3)$  coset picture of $Y^{\!\mathcal M}$ with $\eta_{\ssc \mathcal M\mathcal N} Y^{\!\ssc \mathcal M} Y^{\!\ssc \mathcal N} =+1$. The overlapping coordinate $W$ allows for the description of the two pairs to be put into a single framework as the full $SO(2,4)/SO(1,3)$ coset. Complementary cosets of $SO(1,4)/SO(1,3)$ and $SO(2,3)/SO(1,3)$ in 
\\%%%%%%%%%%%%%%%%%%%%%%
\bea &&
SO(2,4)/SO(1,4) \times SO(1,4)/SO(1,3) = SO(2,4)/SO(1,3)
\sea \qquad
= SO(2,4)/SO(2,3)  \times SO(2,3)/SO(1,3)
\nonumber\eea
are described by $(Y^\mu, Y')$ with $(Y')^2= (Y^{\ssc  4})^2- (Y^{\ssc  5})^2$ giving $\eta_{\mu\nu} Y^{\mu} Y^{\nu}+ (Y')^2 =+1$, and $(Z^\mu, Z')$ with $(Z')^2= (Z^{\ssc  5})^2- (Z^{\ssc  4})^2$ giving $\eta_{\mu\nu} Z^{\mu} Z^{\nu}- (Z')^2 =-1$, respectively. Following the above analysis, this contraction is to be implemented with new parameters  $\bar\lambda^{\mu}={\lambda} \, \omega^{\mu {\ssc 5}}$, $\bar{p}^{\mu}=p \,{\omega}^{\mu {\ssc  4}}$, and $\bar{f}= -\lambda p \,{\omega}^{\ssc  45}$ in the $\lambda, p \to \infty$ limit, using the new  coordinates $\lambda^\mu = \lambda \, Z^{\mu}$, $p^\mu= p \, Y^{\mu}$, and $r = \lambda p \,W$, under the conditions $Z^{\ssc  5} \sim -1$ and $Y^{\!\ssc  4} \sim 1$. We have \\[.1in]
{$\bullet$  $SO(2,4)/SO(1,3) \to H_{\!\ssc R}(1,3)/SO(1,3) $ ---}  
\bea && \!\!\!\!\!\!
\left(\begin{array}{c} \!\!
d\lambda^\mu   \!\! \\    \!\! d(1)   \!\! \\   \!\! dr \\    \!\!  d(1)   \!\! \\ dp^\mu   \!\!
\end{array}\right)  \!\!  = \!\! 
\left(\begin{array}{ccccc}
\omega^\mu_\nu  &  \bar{\lambda}^\mu & \frac{-1}{p^2} \bar{p}^{\mu}  & 0 & 0\\
\!\!  \!\! \frac{-1}{\lambda}  \bar{\lambda}_\nu   & 0 &  \frac{-1}{\lambda p^2} \bar{f}  & 0  &     0\\
\bar{p }_\nu &   \bar{f} &  0  &  \bar{f} & -\bar{\lambda}_\nu\\
 0 & {0} &  \frac{1}{\lambda^2 p} \bar{f} & 0 &  \frac{-1}{p}  \bar{p}_\nu \!\!  \!\! \\
 0& 0  & \frac{-1}{\lambda^2} \bar\lambda^{\mu}  &  \bar{p}^\mu  &\omega^\mu_\nu
\end{array}\right) \!\!\!
\left(\begin{array}{c}
 \lambda^\nu \\ 1 \\ r \\    1 \\ p^\nu 
\end{array}\right) \!\!
= \!\! \left(\begin{array}{c}
\omega^\mu_\nu   \lambda^\nu  + \bar{\lambda}^\mu \\ 0 \\
 \!\!\!\!   \bar{p}_\nu   \lambda^\nu \!\!  -\bar{\lambda}_\nu   p^\nu  \!\! + 2\bar{f} \!\!\!\! \\   0 \\
\omega^\mu_\nu   p^\nu + \bar{p}^\mu\\
\!\! \end{array}\right) ,
\sea
\eea
%$dr= \bar{p}_\nu   \lambda^\nu Y^{\!\ssc  5} + \bar{f} (Y^{\!\ssc  5} 
%     - Z^{\!\ssc  4} ) +\bar{\lambda}_\nu   p^\nu Z^{\!\ssc  4}$\\
assuming $\lambda^{\ssc  4} \to -\lambda$ and $p^{\ssc  5} \to p$. Identifying the $r$ coordinate as $2f$, or taking $r$ as $\theta$ and $\bar\theta=2\bar{f}$ instead [{\em cf.} Eq.(\ref{hr13p})], we have obtained exactly the symmetric description of the $H_{\!\ssc R}(1,3)/SO(1,3)$ coset. Alternatively, we can think of taking $-\frac{1}{p\lambda} J_{\ssc 45}$ as $2 I$ instead of $F$, which therefore naturally yields $\bar\theta I= \omega^{\ssc 45} J_{\ssc 45}$, giving us $\bar\theta = -2 {p\lambda} \omega^{\ssc 45} =2\bar{f}$. The $W$ (or $r$) coordinate is especially introduced to have $r$ bearing the dimensions of ${p\lambda}$; thereby fitting the contracted symmetry with the $F$, or $I$, generator. This analysis actually indicates that using the generator $I$ provides a more natural picture [{\em cf.} the $\hbar=2$ units for quantum mechanics]. It would be good to have an understanding of the $\lambda^{\ssc  4} \to -\lambda$ and $p^{\ssc  5} \to p$ assumption. Thinking about the two contraction parameters as one, this assumption is firstly the statement that the magnitude of $\lambda^{\ssc  4} (= \lambda Z^{\ssc  4})$  and $p^{\ssc  5} (= p Y^{\ssc  5})$ have to go to $\infty$ with the contraction parameter, {\textit{i.e.}} $|Z^{\ssc  4}| \sim |Y^{\ssc  5}| \to 1$. Otherwise, if they go as any other power of the contraction parameter, one would have either $dr=0$, for the two staying finite, or $dr\to\infty$. Neither case can be thought of as a sensible result. The signs are a bit more tricky. Explicitly, we have $dr= \bar{p}_\nu   \lambda^\nu Y^{\ssc  5} + \bar{f} (Y^{\ssc  5} - Z^{\ssc  4} ) +\bar{\lambda}_\nu   p^\nu Z^{\ssc  4}$. Switching both signs hence only changes $r$ to $-r$, which does not change the actual physical picture being described. Taking both going to $\infty$ with the same sign kills the $\bar{f}$ term, which also seems unreasonable. We are not able to, however, say more about this aspect of the coset contraction picture.

%\subsection{\boldmath $SO(2,4)  \to H_{\!\ssc R}(1,3)$ directly}
\subsection{\textnormal{\textbf{Remarks About Physical Dimensions}}}
We explained above how the physics at the level before and after the first 
$\lambda\to\infty$ contraction would lead one to see $\lambda$ as a 
fundamental constant with physical dimensions. This comes with the 
the pair  $\lambda_{\ssc A}$ and $E^{\ssc A}$ getting physical dimensions 
being reciprocal of one another.  Similarly, the $p\to\infty$ contraction, 
whether taken before or after the $\lambda\to\infty$ limit would lead 
to the introduction of another fundamental physical unit: $[p]$. The 
parameters $p^\mu$ and $p^{\ssc 5}$ would have the dimensions of 
$[p]$, while $X_\mu$ and $X_{\ssc 5}$ have that of $[p]^{-1}$. This also
means that $f$ would have the dimensions of $[p][\lambda]$, and $F$ 
would have that of $[p]^{-1}[\lambda]^{-1}$. The latter is obviously 
essentially that of $\hbar(c^{-1})$; however, the single step contraction 
$SO(2,4)  \to H_{\!\ssc R}(1,3)$ would suggest identifying $p$ and 
$\lambda$ -- both would then be $\hbar^{\frac{1}{2}}(c^{-\frac{1}{2}})$. 
$J$ and $\omega$ remain dimensionless. These are the pictures of 
physical dimensions suggested by the contraction analysis at the 
algebra and coset levels. With that said, however, the dynamical 
picture will usually have $J$ given by the orbital angular momentum; 
hence having the units of $\hbar$. A more proper way of expressing 
this fact should actually be that (the orbital part of) 
$J_{\mu\nu}$ is represented by 
$\frac{(c)}{\hbar} (\hat{X}_\mu \hat{E}_\nu -\hat{E}_\mu \hat{X}_\nu)$. 
%From the perspective of the unitary representations, however, it is more natural to take $\hbar(c^{-1})$ as not having units, and especially with the canonical coherent states, actually taking all quantities without physical dimensions.

\section{Contracting the Lorentz Symmetry at the Quantum Level Before Going to the Classical Limit}
We can also consider first contracting $H_{\!\ssc R}(1,3)$ to the 
relativity symmetry of Schr\"odinger quantum physics before going 
to the classical limit. With $K_i= \frac{1}{c} J_{{\ssc 0}i}$, we must 
also set $P_i = \frac{1}{c} E_i$ while keeping $E_{\ssc 0}$ untouched, 
in order to maintain the Galilean commutation relations between 
$K_i$ and $P_i$. Maintaining the 3D Heisenberg commutation relation 
requires putting $G=\frac{1}{c}F$ and keeping $X_i$ unchanged, 
which forces us to set $T =\frac{1}{c}X_{\ssc 0}$. 
Taking these to the $c \to \infty$ limit, we obtain
\bea &&
[J_{ij}, X_k] = -i(\delta_{jk} X_i - \delta_{ik} X_j) \;,
\qquad
[J_{ij}, P_k] = -i(\delta_{jk} P_i - \delta_{ik} P_j) \;,
\sea
[J_{ij}, K_k] = -i(\delta_{jk} K_i - \delta_{ik} K_j) \;,
\qquad
[K_i, K_j] = -\frac{i}{c^2} J_{ij} \to  0 \;,
\nonumber \\ &&
[K_i, H] = -iP_i \;,
\qquad%\qquad%\qquad\qquad
[K_i, P_j] = - \frac{i}{c^2}\delta_{ij}  H  \to  0 \;,
\quad\quad
[X_i,P_j]= i\delta_{ij} G \;,
\nonumber \\ &&
[T , H]= -i G \;,
\quad\quad
[K_i, T ] = -\frac{i}{c^2}X_i  \to 0 \;,
\quad\quad
[K_i, X_j]=  -i\delta_{ij}  T \;,
\eea
where $H\equiv E_{\ssc 0}$.  The set of generators $\{J_{ij}, K_i, P_i,H\}$ 
provides us with the Newtonian/Galilean symmetry of $G(3)$
as a subalgebra. The $\{J_{ij}, X_i, P_i,G\}$ set supplies us with 
a copy of $H_{\!\ssc R}(3)$. The generators $\{J_{ij}, X_i, K_i,T\}$ 
yields another copy of $H_{\!\ssc R}(3)$. As such, we will 
henceforth denote the full symmetry by $H_{\!\ssc G\!H}(3)$. 
Note also that there is an important difference between the 
way the two $H_{\!\ssc R}(3)$ subalgebras sit inside of 
$H_{\!\ssc G\!H}(3)$.  While the $X$-$P$ commutator is a 
central charge for the full algebra, the $X$-$K$ commutator 
is central only within the $H_{\!\ssc R}(3)$ subalgebra it belongs 
to. 
%The really special result that may look quite a surprise is the
%role of the generator $T$ with $[K_i, X_j]=  -\delta_{ij}  T $
%and $[T , H]= -I$.
Furthermore, observe that the $\tilde{G}(3)$  symmetry 
considered in Refs.\cite{066,070} is more akin to the subgroup 
generated by $\{J_{ij}, X_i, P_i,H, G\}$, though within the framework 
presented there the subgroup generated by $\{J_{ij}, K_i, P_i,H, G\}$ 
could serve equally well, assuming $K_i=mX_i$ (as one has for 
a classical particle within the Newtonian framework). The story 
is somewhat more complicated here as we have a nonzero 
$K$-$X$ commutator. That in and of itself actually causes no 
harm in the context of the coset representations we are concerned 
with here. Actually, that suspicious looking commutator can and
will be killed in the classical limit, as shown below.

In order to retrieve the symmetry for Galilean/Newtonian classical physics,  
we can take another further contraction to kill the $X$-$P$ commutator, 
or more accurately to decouple $G$ which requires also killing the $T$-$H$
commutator. Besides, $[K_i, X_j]=  -i\delta_{ij}  T$ looks strange, at least 
for Newtonian physics in which one should have essentially $K_i=m X_i$ 
for a particle of mass $m$. One would like to kill that commutator too. 
This will leave essentially $[K_i, H] = -iP_i$ as the only nonzero commutators 
not involving $J_{ij}$. We will denote the symmetry resulting from this by 
$S_{\!\ssc G}(3)$. One nice way to achieve this is the contraction obtained 
by taking $K_i^c=\frac{1}{k} K_i$,  $P_i^c=\frac{1}{k} P_i$, $X_i^c=\frac{1}{k} X_i$,  
and $T^c=\frac{1}{k} T$ to the $k \to \infty$ limit. We will however 
implement the contraction in a more messy manner with a few different 
parameters. The basic consideration here is to allow more room to cater 
for fixing the units or physical dimensions of all quantities to match 
exactly with the conventional usage in Newtonian physics, as well as 
to reconcile with the alternative contraction sequence below. 
Explicitly, we take
\bea &&
K_i^c=\frac{1}{k} K_i \;,
\qquad
P_i^c=\frac{1}{k_p k} P_i \;, 
\qquad
H^c = \frac{1}{k_p} H \;,
\sea
X_i^c=\frac{1}{k_x} X_i \;,
\qquad
T^c=\frac{1}{k_x} T \;,
\eea
with $k, k_p\to \infty$ simultaneously and also $k_x \to \infty$.
The last limit in itself does not make essential change to the algebra.
The notation with the superscript is to indicate the quantities as
 classical ones compared to the original which are considered
quantum ones. 

The algebra element transition can be written as
\bea &&
{\frac{1}{2}J_{\mu\nu}\omega^{\mu\nu}} + X_\mu \, p^\mu
+ E_{\mu}  \,\lambda^{\mu} + F f
%\nonumber \\ &=&
%{\frac{1}{2}J_{ij}\omega^{ij}} +  K_i \, c\,\omega^{{\ssc 0}i}
%+ X_{i} \bar{\kappa}^i + T c \bar{\kappa}^{\ssc 0}+ P_i c \lambda^i +H \, \lambda^{\ssc 0} + I c f
%\nonumber \\ && \hspace*{-1in}
\sea \longrightarrow 
{\frac{1}{2}J_{ij}\omega^{ij}} + K_i v^i+ X_{i}  p^i  + T e + P_i x^i + H t + G  g
\sea %\hspace*{-1.5in}
\longrightarrow  %S_{\!\ssc G}(3) :
{\frac{1}{2}J_{ij}\omega^{ij}} + K^c_i v^i_c+ X^c_{i}  \, p^i_c  + T^c  e_c + P^c_i x^i_c + H^c t_c + G  g\;,
\eea
where   $v^i = {c} \,\omega^{{\ssc 0}i}$, $e =  {c} \,p^{\ssc 0}$, 
$x^i = {c} \,\lambda^i$, $t = \lambda^{\ssc 0}$, and $g=  {c} f$, 
which are followed by $v^i_c = k \,v^i$, $p^i_c =  k_x \,p^i$, 
$e_c = k_x  e$, $x^i_c = k_p k \,x^i$, and $t_c = k_p t$. 
Again, the subscripts of $_c$ denote the quantities being classical 
ones. With all this understood, it is straightforward to trace the 
contraction of the $H_{\!\ssc R}(1,3)$ cosets discussed above. 
Note that the $ISO(1,3)$ subgroup of $H_{\!\ssc R}(1,3)$ 
to be factored out of the first coset [{\em cf.} Eq(\ref{c13s})] 
is contracted to the copy of $H_{\!\ssc R}(3)$ obtained from 
the set $\{J_{ij}, X_i, K_i,T\}$. We find that 
\vspace{.1in}
\\%%%%%%%%%%%%%%%%%%%%%%%%%%%%%
$\bullet $ $H_{\!\ssc G\!H}(3)/H_{\!\ssc R}(3)$ from $H_{\!\ssc R}(1,3)/ISO(1,3)$  : ---
\bea \label{hg3s}
\left(\begin{array}{c}
dt \\ dx^i \\ dg \\    0
\end{array}\right) =
\left(\begin{array}{cccc}
0 & \frac{1}{c^2} v_j  & 0  & \bar{t} \\
v^i  & \omega^i_j &  0 &  \bar{x}^i \\
-\bar{e}  & \bar{p}_j  &  0 & \bar{g} \\
 {0}   & {0} &  0 & 0
\end{array}\right)
\left(\begin{array}{c}
t \\ x^j  \\ g \\    1
\end{array}\right)
=\left(\begin{array}{c}
\bar{t} \\
v^i t +\omega^i_j  x^j + \bar{x}^i \\ 
  -\bar{e} t + \bar{p}_j   x^j  +\bar{g}\\   0
\end{array}\right) ;
\eea\vspace{.05in}
%********************************
%\bea %\label{hg3s}
%\left(\begin{array}{c}
%dt \\ d\lambda^i \\ df \\    0
%\end{array}\right) =
%\left(\begin{array}{cccc}
%0 & \frac{1}{c} v_j  & 0  & \bar{t} \\
%\frac{1}{c} v^i  & \omega^i_j &  0 &  \bar{\lambda}^i \\
%-{p_0}  & -p_j  &  0 & \bar{f} \\
% {0}   & {0} &  0 & 0
%\end{array}\right)
%\left(\begin{array}{c}
%t \\ \lambda^j  \\ f \\    1
%\end{array}\right)
%=\left(\begin{array}{c}
%\bar{t} \\
%\frac{1}{c} v^i t +\omega^i_j  \lambda^j + \bar{\lambda}^i \\ 
%  -{p_0} t-  p_j   \lambda^j  +\bar{f}\\   0
%\end{array}\right) \;;
%\eea
%%%%%%%%%%%%%%
 {$\bullet$    $H_{\!\ssc G\!H}(3)/H_{\!\ssc R}(3)\to S_{\!\ssc G}(3)/S(3)$  : ---}
\bea \label{sg3s}
\left(\begin{array}{c}
dt_c \\ dx^i_c  \\ dg \\    0
\end{array}\right) =
\left(\begin{array}{cccc}
0 & 0  & 0  & \bar{t}_c \\
v^i_c  & \omega^i_j &  0 &  \bar{x}^i_c  \\
-\frac{1}{k_x k_p}{\bar{e}_c}  & \frac{1}{k_x k_p} \bar{p}_{cj}  &  0 & \bar{g} \\
 {0}   & {0} &  0 & 0
\end{array}\right)
\left(\begin{array}{c}
t_c  \\ x^j_c   \\ g \\    1
\end{array}\right)
=\left(\begin{array}{c}
\bar{t}_c  \\
v^i_c t_c  +\omega^i_j  x^j _c + \bar{x}^i_c  \\ 
  \bar{g}\\   0
\end{array}\right) ;
\eea
%%%%%%%%%%%%%%
$\bullet $ $H_{\!\ssc G\!H}(3)/ISO(3)$ from $H_{\!\ssc R}(1,3)/SO(1,3)$  : ---
\bea \label{hg3p} &&\!\!\!\!\!\!\!\!
\left(\begin{array}{c}
de \\ dp^i \\ dt \\  dx^i \\ dg \\    0
\end{array}\right)  \!\! = \!\!
\left(\begin{array}{cccccc}
 0 &v_j & 0   &  0 & 0 & \bar{e} \\
\frac{1}{c^2}  v^i & \omega^i_j & 0   &  0 & 0 & \bar{p}^i \\
0   &  0 & 0 & \frac{1}{c^2} v_j & 0 & \bar{t} \\
0   &  0 & v^i & \omega^i_j & 0  &  \bar{x}^i   \\
0   &  0 & -\bar{e}   &  \bar{p}_j &  0 &  \bar{g}\\
0   &  0 & {0}   & {0} &  0 & 0
\end{array}\right) \!\!
\left(\begin{array}{c}
e \\ p^j \\ t   \\  x^j \\ {g} \\    1
\end{array}\right) \!\!
= \!\! \left(\begin{array}{c}
v_j p^j + \bar{e}  \\ \omega^i_j p^j + \bar{p}^i \\
 \bar{t} \\ v^i t + \omega^i_j x^j + \bar{x}^i  \\ 
    -\bar{e} t +  \bar{p}_j   x^j  +\bar{g}\\   0
\end{array}\right) ;
\sea
\eea
% only to K and T kills vt, vp, et !!!
%%%%%%%%%%%%%%
 {$\bullet$    $H_{\!\ssc G\!H}(3)/ISO(3)\to S_{\!\ssc G}(3)/ISO(3)$  : ---}
\bea \label{sg3p} &&\!\!\!\!\!\!\!\!\!\!
\left(\begin{array}{c} 
de_c \\ dp^i_c \\ dt_c \\  dx^i_c \\ dg \\    0
\end{array}\right)  \!\! = \!\!
\left(\begin{array}{cccccc}
 0 & \frac{1}{k}v_{cj} & 0   &  0 & 0 & \bar{e}_c \\
0 & \omega^i_j & 0   &  0 & 0 & \bar{p}^i_c \\
0   &  0 & 0 & 0 & 0 & \bar{t}_c \\
0   &  0 &   v^i_c & \omega^i_j & 0  &  \bar{x}^i_c  \\
0   &  0 & -\frac{1}{k_x k_p}{\bar{e}_c}  & \frac{1}{k_x k_p} \bar{p}_{cj}  &  0 &  \bar{g}\\
0   &  0 & {0}   & {0} &  0 & 0
\end{array}\right) \!\!
\left(\begin{array}{c}
e_c \\ p^j_c \\ t_c   \\  x^j_c \\ {g} \\    1
\end{array}\right)\!\!
= \!\! \left(\begin{array}{c}
 \bar{e}_c  \\ \omega^i_j p^j_c + \bar{p}^i_c \\
 \bar{t}_c \\ \!\!  v^i_c t_c + \omega^i_j x^j_c + \bar{x}^i_c \!\!  \\ 
  \bar{g}\\   0
\end{array}\right) .
\sea
\eea

The first thing we want to note regarding the above results is that 
the classical picture of what would be the configuration/physical 
space [Eq.(\ref{sg3s})] and phase space [Eq.(\ref{sg3p})] are very 
good. While we have a relativity symmetry group identification that 
is bigger than the Galilei group, the corresponding cosets are 
essentially trivial extensions of those from the latter. The 
simultaneous existence of $K_i^c$ and $X_i^c$ as now commuting 
generators allows for the standard relation of $K_i^c=mX_i^c$, 
which can be taken as a relation between the representations 
of interest for the otherwise independent generators of the 
background symmetry algebra. The (infinitesimal) momentum 
translations, as given by $\bar{p}_c^i$, can be interpreted as 
merely a consequence of a Galilean boost with 
$\bar{p}_c^i = m v_c^i$ imposed. As said before, one 
expects the latter equation to be retrieved from a Hamiltonian 
equation of motion under the proper setting. We have also 
the extra -- but completely decoupled -- energy ($e_c$) and 
`quantum phase' ($g$) translation symmetries, which are 
irrelevant to the irreducible representations as given by the 
standard Newtonian configuration/physical space of $x^i_c$ 
and momentum space of $p^i_c$. The phase space is the simple 
sum of the two, and consequently a reducible representation. 
The energy translation picture is even there in Newtonian 
physics as the arbitrariness in setting a reference zero point 
for potential energy. This is an incredibly encouraging result, 
indicating that the full scheme envisioned here can make good 
sense from a physical perspective. The symmetry picture 
obtained here for the quantum level, however, needs to be 
considered more carefully, and so this will be addressed below.

\section{Einsteinian/Minkowski Contracted to Galilean/Newtonian Physics}
%%%%
We have briefly addressed the various classical limits of our 
$(1+3)$D picture of a quantum relativity symmetry, and in 
particular the symmetry $S(1,3)$, in Section~2. We take up the 
issue further here, and trace its contraction to 3D classical 
relativity.  Again, the key feature is that $H_{\!\ssc R}(1,3)$
is quite a bit bigger than the usually considered Poincar\'e 
symmetry. Therefore, its classical limit is likely to be also 
somewhat different from the standard Einsteinian relativity. 
The question is whether or not it gives a sensible physical picture 
-- one which includes the latter in some sort of limit. Again, 
we first focus on the coset structures.
 
Firstly, we give the infinitesimal transformation descriptions of the 
two relevant cosets,  namely from Eqs.(\ref{c13s}) and (\ref{c13p}),
under the contraction to $S(1,3)$. The latter is taken here explicitly
as  the $k_x, k_p \to \infty$ limit of $X_\mu^c =\frac{1}{k_x} X_\mu$ 
and $E_\mu^c =\frac{1}{k_p} E_\mu$. The results are simply 
given by \\[.1in]
%%%
{$\bullet$  $S(1,3)/ISO(1,3)$ from $H_{\!\ssc R}(1,3)/ISO(1,3)$: ---}  
\bea\label{s13s}
\left(\begin{array}{c}
d\lambda^\mu_c \\  df \\    0
\end{array}\right) =
\left(\begin{array}{ccc}
\omega^\mu_\nu & 0 &  \bar{\lambda}^\mu_c  \\
       \frac{1}{k_x k_p} \bar{p}_{c\nu}   &  0 &   \bar{f}\\
 {0}   & {0} &  0
\end{array}\right)
\left(\begin{array}{c}
  \lambda^\nu_c \\ {f} \\    1
\end{array}\right)
=\left(\begin{array}{c}
\omega^\mu_\nu   \lambda^\nu_c  +   \bar{\lambda}^\mu_c   \\ 
   \bar{f}\\   0
\end{array}\right) \;,
\eea\vspace{.05in}
%%%%%%%%%%%%%%%%%
{$\bullet$  $S(1,3)/SO(1,3)$  from $H_{\!\ssc R}(1,3)/SO(1,3)$: ---}  
\bea \label{s13p}
\left(\begin{array}{c}
dp^\mu_c   \\ d\lambda^\mu_c \\ df \\    0
\end{array}\right) =
\left(\begin{array}{cccc}
 \omega^\mu_\nu &  0& 0  &  \bar{p}^\mu_c \\
0  & \omega^\mu_\nu &  0 &  \bar{\lambda}^\mu_c \\
 0 &  \frac{1}{k_x k_p} \bar{p}_{c\nu}  &  0 & \bar{f} \\
 {0}   & {0} &  0 & 0
\end{array}\right)
\left(\begin{array}{c}
p^\nu_c  \\ \lambda^\nu_c   \\ f \\    1
\end{array}\right)
=\left(\begin{array}{c}
\omega^\mu_\nu   p^\nu_c  + \bar{p}^\mu_c \\
\omega^\mu_\nu   \lambda^\nu_c   + \bar{\lambda}^\mu_c  \\ 
 \bar{f} \\   0
\end{array}\right) \;,
\eea
where ${\lambda}^\mu_c  = k_p {\lambda}^\mu$ and $p^\mu_c= k_x p^\mu$.
Note that these results are exactly the same, apart from a normalization 
of the decoupled `quantum phase' (as given by $f$ here), as if we had 
applied the contraction to the forms of the cosets as given in Sec.II 
instead. We have Minkowski spacetime arising as 
$S(1,3)/ISO(1,3) \sim ISO(1,3)/SO(1,3)$, here described by 
four `time' coordinates, and a matching phase space with four 
additional momentum coordinates. As discussed somewhat 
in the introductory section, these momentum translations 
are beyond the standard Einsteinian formulation. Note that here 
they are transformations independent from, and in addition to,
 the Lorentz boosts, as described infinitesimally by the 
$\omega^\mu_\nu$. Thinking about the $E_\mu$ generators 
and the corresponding $p^\mu$ parameters as describing the 
energy-momentum four-vector as observables, having their 
components transforming as a Lorentz four-vector is of 
course an actual necessity, so long as all $p^\mu$, for 
example, are to be included as phase space coordinates.

The next contraction to consider is again taking the Lorentz 
boosts to the Galilean boosts, as in the last section. We take  
$K_i^c= \frac{1}{c k} J_{{\ssc 0}i}$, $P^c_i = \frac{1}{c k} E^c_i$,
and $T^c = \frac{1}{c} X^c_{\ssc 0}$ to the $c, k \to \infty$ limit, 
obtaining
\bea &&
[J_{ij}, K_k^c] = -(\delta_{jk} K_i^c - \delta_{ik} K_j^c) \;,
\qquad
%\nonumber \\ &&
[K_i^c, K_j^c] = -\frac{1}{c^2 k^2} J_{ij} \to  0 \;,
\nonumber \\ &&
[K_i^c, H^c] = -P^c_i \;,
\qquad\qquad\qquad\qquad
[K_i^c, P^c_j] = - \frac{1}{c^2 k^2}\delta_{ij}  H^c  \to  0 \;,
\nonumber \\ &&
[J_{ij}, P^c_k] = -(\delta_{jk} P^c_i - \delta_{ik} P^c_j) \;,
\qquad\quad
[J_{ij}, H^c] = 0 \;,
\qquad\quad
[P^c_i, H^c] = 0 \;,
\sea
[K_i^c, T^c ] = -\frac{1}{c^2 k} X_i^c  \to 0 \;,
\qquad\quad\quad\;
[K_i^c, X_j^c]=  -\frac{1}{k} \delta_{ij}  T^c   \to 0\;,
\eea
where we have introduced $H^c\equiv E^c_{\ssc 0}$.  
This result,  is exactly the same as the result given in the last section,
 for $S_{\!\ssc G}(3)$. The non-minimal forms we have been writing
the contractions of $H_{\!\ssc G\!H}(3) \to S_{\!\ssc G}(3)$ above 
and the $S(1,3) \to S_{\!\ssc G}(3)$ here give room to accommodate
that exact matching of final results for the two alternative contraction
sequences from $H_{\!\ssc R}(1,3)$.  

The cosets for the Newtonian configuration/physical space(-time) 
and phase space, exactly as given in the last section, are also to be 
obtained from this alternative line of contractions, which are 
given by \\[.1in]
%%%%%%%%%%%%%
 {$\bullet$   $S(1,3)/ISO(1,3) \to S_{\!\ssc G}(3)/S(3)$ :  ---}
\bea 
\left(\begin{array}{c}
dt_c \\  dx^i_c \\ df \\    0
\end{array}\right) =
\left(\begin{array}{cccc}
0 & \frac{1}{c^2 k^2} v_{c j} & 0 & \bar{t}_c \\
 v^i_c & \omega^i_j & 0  &  \bar{x}^i_c   \\
0   &  0 &  0 &  \bar{f}\\
 {0}   & {0} &  0 & 0
\end{array}\right)
\left(\begin{array}{c}
t_c   \\  x^j_c \\ {f} \\    1
\end{array}\right)
=\left(\begin{array}{c}
\bar{t}_c \\ v^i_c t_c + \omega^i_j x^j_c + \bar{x}^i_c  \\ 
  \bar{f}\\   0
\end{array}\right) \;,
\eea
{$\bullet$   $S(1,3)/SO(1,3) \to S_{\!\ssc G}(3)/ISO(3)$  : ---}
\bea \label{sg3p} && \!\! \!\! \!\! \!\! 
\left(\begin{array}{c}
de_c \\ dp^i_c \\ dt_c \\  dx^i_c \\ df \\    0
\end{array}\right)  \!\!  = \!\! 
\left(\begin{array}{cccccc}
 0 & \frac{1}{k} v_{c j} & 0   &  0 & 0 & \bar{e}_c \\
 \frac{1}{c^2 k} v^i_c & \omega^i_j & 0   &  0 & 0 & \bar{p}^i_c \\
0   &  0 & 0 & \frac{1}{c^2 k^2} v_{c j} & 0 & \bar{t}_c \\
0   &  0 & v^i_c & \omega^i_j & 0  &  \bar{x}^i_c   \\
0   &  0 & 0  & 0  &  0 &  \bar{f}\\
0   &  0 & {0}   & {0} &  0 & 0
\end{array}\right) \!\! 
\left(\begin{array}{c}
e_c \\ p^j_c \\ t_c   \\  x^j_c \\ {f} \\    1
\end{array}\right) \!\! 
= \!\!  \left(\begin{array}{c}
 \bar{e}_c  \\ \omega^i_j p^j_c + \bar{p}^i_c \\
 \bar{t}_c \\   v^i_c t_c + \omega^i_j x^j_c + \bar{x}^i_c  \\ 
  \bar{f}\\   0
\end{array}\right) ,
\sea
\eea
where $v^i_c = c k \omega^{{\ssc 0}i}$, $x^i_c=c k \lambda_c^i$, 
and $e_c = c p_c^{\ssc 0}$, with $t_c \equiv \lambda_c^{\ssc 0}$ .

\section{On The Physical Dimensions of Quantities and the Nature of $\hbar$}
We would like to look into the issue of physical dimensions in some 
more detail. We have discussed how the contraction processes, as well 
as the studying of theories on the lower levels of the contraction sequences, 
suggest the introduction of (relative) physical dimensions to various 
quantities in terms of a `natural' choice of different physical units; units 
which would {\em not} naturally be used in the more fundamental theory. 
The relationship between the units chosen by humans and the truly 
natural (numerical) representations of such `quantities' in the 
mathematical structure lurking beneath gives rise to fundamental 
constants in physics such as $c$, $\hbar$, and $G$. For example, 
taking the Galilean approximation of Lorentz symmetry suggests 
space and time are independent; hence to be measured in different 
units. With Lorentz symmetry, $\frac{1}{c^2}$ is just a structural 
constant of the $SO(1,3)$ symmetry algebra, which is stable against 
deformation \cite{067}, meaning any nonzero value(s) of the 
structural constants $\frac{1}{c^2}$ give the same symmetry, and 
the natural choice is $c$ being unity and dimensionless. This 
corresponds to a nontrivial, fixed value in, say, meters per second.
We have also explicitly traced the physical dimensions of quantities 
through each of the steps of various contractions in Section~3. The latter
 illuminates an idea that can be easily applied to all of the other cases 
discussed above.  

Let us try to see what we can learn from this coupled with our practical 
usage of physical dimensions. Firstly, we tabulate all of the quantities 
with physical dimensions obtainable from tracing through all the above 
contraction analysis. In Table 1, we present the results for each relativity 
symmetry level -- first for the coset coordinates and the parameters 
of (infinitesimal) transformations, followed by the symmetry generators. 
Note that the former parameters serve as coordinates of the group 
manifolds, coordinates of the cost spaces, and parameters each for
a one-parameter group of transformations generated by a particular
symmetry generator, while the latter correspond also to physical
observables or conserved quantities under the transformations.
Moreover, we have a cross matching between space-time and
energy-momentum parts; the momentum observables $P_\mu$ are 
generators of translations in positions $x^\mu$ the observables of 
which as $X_\mu$ generates translations of $p^\mu$. We want to be 
able to identify at least the physical dimensions of $P_\mu$ and $p^\mu$ 
as well as $X_\mu$ and $x^\mu$, and similarly for the corresponding
quantities at various levels of the contractions from $H_{\!\ssc R}(1,3)$
onwards, {\em e.g.} $E^c_\mu$ and $\lambda^\mu_c$ or $T$ and $t$.
In fact, we have arranged in those columns like $p^i$ with $P_i$
instead of $x^i$ with $P_i$ as in $\omega^{\mu\nu}$ with 
$J_{\mu\nu}$. To achieve that identification of physical 
dimensions, one needs to rescale all generators by a constant
parameter with nontrivial dimension. That is the common
physicists' convention, like putting the generators of rotations
as angular momentum with its physical dimension. We apply 
exactly that to give results in the last row under each symmetry.
It works perfectly except in the case of $S_{\!\ssc G}(3)$ 
in which we have to further take the contraction parameter $k$
to have no physical dimension.  With that assumed,  we can see 
then for the quantities under $S_{\!\ssc G}(3)$ to be those taken 
in Newtonian physics we have $[\lambda ][k_x] = [T]$ and
$[p][k_p] = [M] [L] [T]^{-1}$ where $c$ has been taken as the
speed of light hence $[c]=[L][T]^{-1}$. Moreover, those are the
three basic  units or independent fundamental constants. In fact,
we usually describe quantum mechanics without and with Lorentz 
symmetry (as in the so-called relativistic quantum mechanics) 
using physical quantities in the same units as their classical limits.
The quantum theories are described with relativity symmetries 
$H_{\!\ssc G\!H}(3)$ and $H_{\!\ssc R}(1,3)$, respectively, here;
and  $S_{\!\ssc G}(3)$  and $S(1,3)$ are the classical limits.
That would correspond to having no physical dimensions for
$k_x$ and $k_p$, hence $\lambda$ and $p$ as the invariant
time and momentum, respectively. $\lambda cp$ should then
be $\hbar$. That is the basic picture discussed along with the
$SO(2,4) \to H_{\!\ssc R}(1,3)$ contractions above.  

A very interesting point to note from all the analysis about the
notion of physical dimensions is that the Planck constant
$\hbar$, though usually taken to be characterizing the quantum
nature of things actually has its origin mostly from the higher level
contraction. Though taking the naive $\hbar \to 0$ limit trivializes 
the Heisenberg commutation relation and decouples the central 
charge generator for the quantum phase, that should not be taken
as the relativity symmetry contraction to give classical limits. This 
result is in line with our analysis in Ref.\cite{070}. Unlike $c$, which 
is the contraction parameter introduced for contraction from 
Lorentz symmetry to Galilean, $\hbar$, or its inverse, plays no such 
role between quantum and classical physics. Physics with Lorentz
symmetry should be described in $c=1$ units with space and time
described in the same way, which is well justified from the
contraction analysis here. Taking $\hbar=1$ units for quantum
physics however cannot be justified in parallel.

Let us look at the physical dimension issue in relation to $\hbar$ 
from another point of view. A single step contraction from $SO(2,4)$ 
yields $H_{\!\ssc R}(1,3)$  with one parameter which would 
essentially be $\sqrt{\hbar}$: $\lambda=p=\sqrt{\hbar}$~\footnote{
%%%%%%~\footnote{
The standard $\hbar$ dimension is that of $\lambda pc$; hence giving 
the correct choice as $\lambda=p=\sqrt{\frac{\hbar}{c}}$. Here, we are 
neglecting $c$, which should be taken to be trivial at this level. All of 
the $\hbar$'s here should actually be $\frac{\hbar}{c}$. 
The exact dimensions -- including the $c$ factors -- are given in Table~2.}.
%%%%%%
The resulting structure has a more symmetric role for $E_\mu$ and 
$X_\mu$, and this corresponds to our natural choice of phase space 
coset, as used in Section~2 [{\em cf.} Eq.(\ref{hr13p})], and similarly in 
Refs.\cite{066,070} where the focus is really just the $H(1,3)$, or $H(3)$, 
subgroup. The contraction parameter $\sqrt{\hbar}$ is then introduced 
to get an approximation to the physics of the otherwise $SO(2,4)$ 
relativity symmetry. As such, the approximation is the 
$\hbar\to\infty$ limit! This is the true parallel of 
$\sqrt{\hbar}$ to $c$, based on the results here. 

More explicitly, one takes $E_\mu=-\frac{1}{\sqrt{\hbar}} J_{\mu\ssc 5}$ 
and $X_\mu=\frac{1}{\sqrt{\hbar}} J_{\mu\ssc 4}$ to the limit where 
$[E_\mu, E_\nu] =- [X_\mu, X_\nu] =-\frac{i}{{\hbar}} J_{\mu,\nu} \to 0$. 
Following the contraction notion naively, one would expect 
that the very small quantum $\hbar$ is really to be taken as 
a big parameter, with dimensions much smaller  compared 
to the $SO(2,4)$ physics of noncommuting $E_\mu$ and 
noncommuting $X_\mu$. At the Galilean level, we do not see 
the invariant speed $c$ among the structural constants of the 
relativity symmetry (or otherwise), but we have physics with 
the dimension $[c]$ (or equivalently, different physical 
dimensions for time and distance); at the usual quantum level, 
as in $H_{\!\ssc R}(1,3)$ [or $H_{\!\ssc G\!H}(3)$], we do not 
see $\hbar$ in the the part of the symmetry description, but 
have a notion of the physical dimension of $[\hbar]$. The 
explicit $\hbar$ in the Heisenberg commutation relation
$[\hat{X},\hat{P}]=i\hbar$ should be taken only at the level
of the representation of the symmetry, not that of the
symmetry of the Heisenberg algebra. The central charge
generator has the physical dimension of $\hbar$ and is
represented by $\hbar$ times the identity operator on the
Hilbert space. $c$ being an invariant parameter is an issue of 
Lorentz symmetry; the physics of the $SO(2,4)$ symmetry only 
is to reveal $\hbar$ as an invariant, like under (quantum) reference 
frame transformations. Looking at things from this perspective, 
this does not seem to be unreasonable at all, though it is saying 
that our earlier thinking about the role of $\hbar$ was quite 
wrong. That is one key lesson here.

Beyond the coset level, we have to look at the unitary 
representations arising from coherent state constructions, 
which again suggests the natural choice of using identical 
units for $E_\mu$ and $X_\mu$. Note that the exact nature 
of such formulations at the $(1+3)$D and 3D relativity 
symmetry levels are not the same. At the 3D level of 
$H_{\!\ssc G\!H}(3)$, this means identical units for 
$P_i$ and $X_i$, or equivalently, for $p^i$ and $x^i$. Identical 
units for $E_\mu$ and $X_\mu$ would yield $P_i$ and $X_i$ 
with units differing by a factor of $c$, upon taking the 
Lorentz to Galilean contraction, as shown explicitly in Table~2.

The classical pictures, as presented in Table~2, deserve some 
further attention. By not having different dimensions for the 
`length' and `momentum' at the quantum level (explicitly 
$H_{\!\ssc R}(1,3)$ here), we must introduce that splitting at 
the classical level to retrieve the usual pattern of physical 
dimensions. This is to be achieved by having different nontrivial
physical dimensions for $k_x$ and $k_p$. Not doing this would 
have, for example, kept $X_{\!\mu}^c$ and $E_{\!\mu}^c$ of
$S(1,3)$ as having the same physical dimensions. In fact, we need 
$[k_x]= [k_p]^{-1}$ to have the exact matching with our familiar
usage of classical physical units. If such a two parameter 
contraction sounds odd, one can certainly implement it in two 
separate steps. Adopting this convention, we again have a story
fully consistent with known, practical physics. Note that from
the pure theoretical perspective of successive approximations
to the fundamental physics of $SO(2,4)$ through the contraction
the choice of physical units are not optimal. The practical choice
of our system of physical units is simply contingent on the
human cultural history.

\section{Concluding Remarks}
%%%
In the preceding sections we first put together the relevant 
relativity symmetry contractions and the resulting contractions 
of the relevant coset space representations, starting from the 
$SO(2,4)$ symmetry. The coset space representations are 
what can be called the (configuration) space coset and the 
phase space coset at each level, which in the classical cases 
give pictures of space and phase space for a single particle 
system (or the center of mass of a system of particles). We 
have considered alternative contraction sequences, such as 
taking the Lorentz symmetry to the Galilean limit first before 
going from the quantum case to that of classical physics  
and the other way round -- achieving virtually the same 
result. We essentially recover Newtonian space-time and 
phase space, as well as Minkowski spacetime. The $(1+3)$D 
relativity symmetry has a phase space picture with full 
Minkowski energy-momentum four-vector coordinates 
admitting independent translations in all directions. Earlier 
analysis indicates that, at least for Hamiltonian evolution 
as generated by the square of the energy-momentum 
four-vector, one recovers particle dynamics of Einsteinian 
special relativity with proper time (or rather proper 
time/rest mass) as the evolution parameter. All of this 
can be considered as preliminary success of the scheme 
being advocated for here, which now includes all known 
particle dynamical pictures with their identified relativity 
symmetries. This is the key message of this article.

As discussed in the introduction, the relativity symmetry 
perspective requires going beyond the Poincar\'e symmetry 
and its stabilization. Formulating physics consistent with all 
available experimental results within this scheme may 
already be quite a challenge. The more exciting prospect 
of obtaining new predictions is even more interesting. 
A key point to be made here is that the proper interpretation 
of theories of this kind is also likely to require adjustments 
to our understanding of existing physical concepts beyond 
our old frameworks. One example we have discussed a bit 
above is the version of quantum mechanics with the 
$H_{\!\ssc R}(1,3)$ symmetry. The basic features of said 
quantum theory would be like those found in covariant 
quantum mechanics, which has been studied by various 
authors before while being mostly neglected by other 
physicists. Extending a $\phi(x^i)$ wavefunction to one 
of the $\phi(x^\mu)$ wavefunctions, and considering its 
proper time evolution, sounds like a very natural way to 
put quantum mechanics within the framework of Einsteinian 
special relativity. There has actually been a long history, 
comprised of many diverse efforts, along this line, which 
also brought up key notions such as mass indefiniteness, 
the introduction and interpretation of antiparticles, and 
the direction of time (etc.), which are to be addressed 
below. As such, we do not even consider it of much 
interest to the readers to cite here more references 
pertaining to these issues. On the other hand, though, 
a particularly noteworthy reference comes from 
Feynman's work on quantum electrodynamics \cite{F}. 
The master went beyond everybody, actually taking 
the Klein-Gordon equation exactly as the 
$\sigma$-independent equation (again 
$\sigma=\frac{\tau}{m}$) of the $\sigma$-evolution 
Schr\"odinger equation in covariant quantum 
mechanics, and discussed the $\frac{dt}{d\sigma}<0$ 
case in connection with the notion of antiparticles
 \cite{ap}. Readers interested in more details 
regarding this point are suggested to consult 
Refs.\cite{036,037}, which present analyses in 
that direction based on a setting that is somewhat 
different from, but very compatible with, the one 
presented here. A key point is that -- like what 
lies behind the wisdom of St\"uckelberg-Feynman
 -- all theoretical results presented there should be 
(re-)interpreted in laboratory terms, based on a 
coordinate time $t$ with its forward-increasing 
direction, at least to the extent that a classical time 
idea is involved at all. For example, things `evolving' 
backwards in time is to be interpreted in our 
forward-increasing time somewhat differently -- 
in exactly the same vein as thinking of a particle 
moving backward in time as an antiparticle 
moving forward in time. 

We want to leave this subject matter until we have a full dynamical 
analysis of the $H_{\ssc R}(1,3)$ theory and its contraction limits, 
except for one final point: the question of the interpretation of the 
$\phi(x^\mu)$  wavefunction. A Born probability interpretation may 
be tricky. It has considered to be a consistent framework though \cite{Fi}.
We have an alternative picture to advocate. Our perspective 
is that quantum mechanics is about quantum models of spacetime to 
which the classical models provide only an approximation. These 
quantum models, like the example discussed in  Ref.\cite{066}, are 
not finite-dimensional, real number geometries. The (projective) 
Hilbert space of a quantum system provides one with a 
real-number-geometric description of the otherwise noncommutative 
geometry. A wavefunction as a description of a vector in the Hilbert 
space is really the infinite number of coordinates under a fixed 
choice of coordinate frame, where the basis is being provided 
by $\left|x^\mu\rra$ in  this case. A quantum state has a 
completely fixed position (in the quantum spacetime) without 
an uncertainty. However, such a position is to be described by 
some noncommutative values instead of real number values, 
or equivalently an infinite number of the latter. How single 
classical-physics-like real number values of a (repeated) 
von-Neumann measurement, and the probability distribution 
of such results, is obtained is only a matter of the kind of 
measurements being performed. This can be considered 
quite well-explained by decoherence theory, at least for 
standard quantum mechanics. Of course, how to better 
understand the nature of these noncommutative coordinate 
values, especially for the case of the physical time variable 
$\hat{T}$, is a key challenge.

Among all of the coset results presented here, the one whose details 
are less obvious -- which apparently would provide quite a challenge 
in formulating a dynamical theory -- is actually the case of 3D quantum 
relativity symmetry $H_{\ssc G\!H}(3)$ [{\em cf.} Eqs.(\ref{hg3s}) and 
(\ref{hg3p})]. Looking at it from the formulation of quantum mechanics 
as discussed in Ref.\cite{070}, however, we see that this does not actually 
present much of problem. We still have $H(3)$ as an invariant subgroup. 
The cosets still possess an absolute Newtonian time, at least if we write 
quantum mechanics based on the first coset, {\em i.e.} with the 
$\phi(x^\mu)$  wavefunction. The only somewhat nontrivial issue is 
the quantum phase contribution from the energy-time product. So 
long as we do not consider energy translations, the extra contribution 
drops out; hence, the usual formulation works perfectly, at least as 
a special case. While a canonical coherent state formulation from the 
phase space coset looks somewhat complicated, we know the physical 
story has to be the same as in the other case, as the irreducible unitary 
representation on the Hilbert space for the $H(3)$ part is essentially 
unique.  What is promising is that, again, a unitary Hilbert space 
representation of $H_{\!\ssc R}(1,3)$, with the corresponding extension 
to its group $C^*$-algebra (at least for the case of the canonical coherent 
states), should give a fully dynamical picture of the theory. The case has 
been fully elaborated for the $H_{\!\ssc R} (3)$ setting in Ref.\cite{070}.
The various contractions of which would give the corresponding 
dynamical descriptions at the $H_{\!\ssc G\!H}(3)$, $S(1,3)$, and 
$S_{\!\ssc G}(3)$ levels, which should agree with known physics. The 
coset-level story appears promising enough, as illustrated here. What 
else can be learned from such a fully dynamical analysis is the exciting 
task at hand, on which we hope to be able to report soon. The grand 
game plan is, of course, to push back up to the highest level of 
$SO(2,4)$ relativity and formulate its dynamical picture.

So, to conclude, the analysis presented above of contractions of coset 
representations indicates no inconsistency with established theories, 
which should be considered as successfully retrieved from the appropriate 
limits and special cases. More has to be learned and some 
technically-detailed challenges remain to be surmounted, from 
which we may discover new features about Nature.

 %%%%%%%%%%%%%%%%%%

\vspace*{.2in}
\noindent{\bf Acknowledgements \ }
The authors are partially supported by research grants 
number 105-2112-M-008-017 and  106-2112-M-008-008
of the MOST of Taiwan.

\newpage
\begin{sidewaystable}[htb]
%\footnotesize
%\begin{table}
\caption{Table on Physical Dimensions of Quantities. 
($[pc]=[p][c]$ and $[\lambda kc]^{-1}=[\lambda]^{-1}[k]^{-1}[c]^{-1}$,  etc.) 
(Note that we need $[k]$ to be 1, {\em i.e.} $k$ as dimensionless).}

\begin{center}
\begin{tabular}{|c|c|c|c|c|c|c|c|}    \hline\hline 
$SO(2,4)$ 			&  \multicolumn{7}{|c|}{$\omega^{\!\ssc \mathcal{M\!N}}$}	 	\\
		& \multicolumn{7}{|c|}{$J_{\!\ssc \mathcal{M\!N}}$} \\ \hline
$ISO(1,4)$			&  \multicolumn{2}{|c|}{$\lambda^{\mu}$ -- $[\lambda]$}   & \multicolumn{4}{|c|}{$\omega^{\ssc A\!B}$} & {$\lambda^{\ssc 4}$ -- $[\lambda]$} \\
		&  \multicolumn{2}{|c|}{$E_{\!\mu}$ -- $[\lambda]^{-1}$}   & \multicolumn{4}{|c|}{$J_{\!\ssc A\!B}$}  & {$E_{\ssc 4}$ -- $[\lambda]^{-1}$}\\ \hline
$ISO(2,3)$			&  \multicolumn{4}{|c|}{$\omega^{\mu\ssc 5}$, $\omega^{\mu\nu}$ }	& \multicolumn{2}{|c|}{$p^{\mu}$  -- $[p]$}	 & {$p^{\ssc 5}$  -- $[p]$}  \\
 & 	\multicolumn{4}{|c|}{$J_{\mu\ssc 5}$, $J_{\mu\nu}$ }	& \multicolumn{2}{|c|}{$X_{\mu}$ -- $[p]^{-1}$ }	 & {$X_{\ssc 5}$ -- $[p]^{-1}$} \\  \hline
%%%%%
$H_{\!\ssc R}(1,3)$	&	\multicolumn{2}{|c|}{$\lambda^\mu$ -- $[\lambda]$}  & \multicolumn{2}{|c|}{$\omega^{\mu\nu}$}	 			& \multicolumn{2}{|c|}{$p^\mu$ -- $[p]$}  &	$f$ -- $[\lambda p]$	\\
		&  \multicolumn{2}{|c|}{$X_\mu$ -- $[p]^{-1}$}    &  \multicolumn{2}{|c|}{$J_{\mu\nu}$} &  \multicolumn{2}{|c|}{$E_\mu$ -- $[\lambda]^{-1}$}  & $F$ -- $[\lambda p]^{-1}$ \\ 
$\times  [\lambda p]$		&  \multicolumn{2}{|c|}{$X_\mu$ -- $[\lambda]$}    &  \multicolumn{2}{|c|}{$J_{\mu\nu}$ -- $[\lambda p]$ } &  \multicolumn{2}{|c|}{$E_\mu$ -- $[p]$}  & $F$ -- \\  \hline
%%%%%%%
$H_{\!\ssc G\!H}(3)$	&	$t$ -- {$[\lambda ]$}   & $x^i$ --	{$[\lambda c]$} 	& $\omega^{ij}$	 & $v^i$ -- $[c]$  & $p^i$ --  {$[p]$} & $e$ -- {$[pc]$}    &  $g$ -- $[\lambda pc]$\\ 
  & $T$ -- $[pc]^{-1}$  & $X_{\! i}$ -- $[p]^{-1}$    & $J_{ij}$   & $K_i$ --  $[c]^{-1}$   & $P_i$  -- $[\lambda c]^{-1}$  & {$H$  -- $[\lambda]^{-1}$} &  $G$ -- $[\lambda pc]^{-1}$  \\ 	
 $\times  [\lambda pc]$ & $T$ -- {$[\lambda]$}   & $X_{\! i}$ -- {$[\lambda c]$}     & $J_{ij}$ -- $[\lambda pc]$   & $K_i$ --  $[\lambda p]$   & $P_i$  -- $[p]$  & {$H$  -- $[pc]$} &  $G$ --   \\ 	  \hline
$S_{\ssc G}(3)$	&	$t_c$ -- {$[\lambda k_p]$} 	& $x^i_c$ --	{$[\lambda c k k_p]$}    &  $\omega^{ij}$	& $v_c^i$ -- $[c k]$ & $p^i_c$ --  {$[p k_x]$} 	 & $e_c$ -- {$[pc k_x]$}  &  \\  	 
  &  $T^c$  -- $[pck_x]^{-1}$ &  $X_{\!i}^{\!c}$  -- $[pk_x]^{-1}$  & $J_{ij}$ & $K_i^c$ --  $[ck]^{-1}$ & $P^c_i$  -- $[\lambda ckk_p]^{-1}$ &{$H^c$  -- $[\lambda k_p]^{-1}$} &  \\    
$\times  [\lambda pck  k_xk_p]$ &  $T^c$  -- $[\lambda kk_p]$ &  $X_{\!i}^{\!c}$  -- $[\lambda ckk_p]$  & $J_{ij}$ -- $[\lambda pck  k_xk_p]$ & $K_i^c$ --  $[\lambda pk_xk_p]$ & $P^c_i$  -- $[pk_x]$ & {$H$  -- $[pc kk_x]$} &  \\  \hline
$S(1,3)$				&	\multicolumn{2}{|c|}{$\lambda^\mu_c$  -- {$[\lambda k_p]$}} & \multicolumn{2}{|c|}{$\omega^{\mu\nu}$}  	&\multicolumn{2}{|c|}{$p^\mu_c$ --  {$[pk_x]$}}    &	$f$ -- $[\lambda p]$ 	\\
		&  \multicolumn{2}{|c|}{$X^c_\mu$ -- $[pk_x]^{-1}$}  &  \multicolumn{2}{|c|}{$J_{\!\mu\nu}$}   &
\multicolumn{2}{|c|}{$E^c_\mu$ -- $[\lambda k_p]^{-1}$}  & $F$ -- $[\lambda p]^{-1}$ \\  
$\times  [\lambda pk_xk_p]$		&  \multicolumn{2}{|c|}{$X^c_\mu$ -- $[\lambda][k_p]$}  &  \multicolumn{2}{|c|}{$J_{\!\mu\nu}$ -- $[\lambda pk_x k_p]$}   &
\multicolumn{2}{|c|}{$E^c_\mu$ -- $[p k_x]$}  & $F$ -- $[k_xk_p]$ \\ \hline
%$S_{\ssc G}(3)$	&	$t_c$ -- {$[\lambda][k]$} & $x^i_c$ --	{$[\lambda][c][k]$}    &  $\omega^{ij}$	& $v^i$ -- $[c]$ & $p^i_c$ --  {$[p][k]$} 	 & $e_c$ -- {$[p][c][k]$}  &  $f$ -- $[\lambda][p]$\\ 
 % &  $X_{\!\ssc 0}$ -- $[p]^{-1}$  & $X_{\! i}$ -- $[p]^{-1}$  & $J_{ij}$ & $K_i$ --  $[c]^{-1}$ & $P^c_i$  -- $[\lambda ck]^{-1}$ &{$H^c$  -- $[\lambda k]^{-1}$} & $F$ -- $[\lambda p]^{-1}$ \\
 %   $\times  [\lambda] [p][c][k]^2$ &  $X_{\!\ssc 0}$ -- $[\lambda][c][k]^2$  & $X_{\! i}$ -- $[\lambda][c][k]^2$  & $J_{ij}$ -- $[\lambda] [p][c][k]^2$ & $K_i$ -- $[\lambda] [p][k]^2$  & $P^c_i$  -- $[p][k]$ &{$H^c$  -- $[p][c][k]$} & $F$ -- $[k]^2$ \\
\hline\hline
\end{tabular}
\end{center} 
\end{sidewaystable}
%\end{table}

\newpage
%\begin{table}
\begin{sidewaystable}[htb]
%\footnotesize
\caption{Table on Physical Dimensions of Quantities starting with one-step
contraction to $H_{\!\ssc R}(1,3)$, equivalent to 
$[\lambda]=[p]=[\hbar]^{\frac{1}{2}}[c]^{-\frac{1}{2}}$. 
(Note that we need $[k]$ to be 1, {\em i.e.} $k$ as dimensionless).}

\begin{center}
\resizebox{1.15\textwidth}{!}{
\begin{tabular}{|c|c|c|c|c|c|c|c|}    \hline\hline
$SO(2,4)$ 			&  \multicolumn{7}{|c|}{$\omega^{\!\ssc \mathcal{M\!N}}$}	 	\\
		& \multicolumn{7}{|c|}{$J_{\!\ssc \mathcal{M\!N}}$} \\ \hline
$H_{\!\ssc R}(1,3)$	&	\multicolumn{2}{|c|}{$\lambda^\mu$ -- $[\hbar]^{\frac{1}{2}}[c]^{-\frac{1}{2}}$}  & \multicolumn{2}{|c|}{$\omega^{\mu\nu}$}	 			& \multicolumn{2}{|c|}{$p^\mu$ -- $[\hbar]^{\frac{1}{2}}[c]^{-\frac{1}{2}}$ }  &	$f$ -- $[\hbar][c]^{-1}$	\\
		&  \multicolumn{2}{|c|}{$X_\mu$ -- $[\hbar]^{-\frac{1}{2}}[c]^{\frac{1}{2}}$}    &  \multicolumn{2}{|c|}{$J_{\!\mu\nu}$} &  \multicolumn{2}{|c|}{$E_\mu$ -- $[\hbar]^{-\frac{1}{2}}[c]^{\frac{1}{2}}$}  & $F$ -- $[\hbar]^{-1}[c]$ \\ 
$\times  [\hbar][c]^{-1}$		&  \multicolumn{2}{|c|}{$X_\mu$ -- $[\hbar]^{\frac{1}{2}}[c]^{-\frac{1}{2}}$}    &  \multicolumn{2}{|c|}{$J_{\!\mu\nu}$ -- $[\hbar][c]^{-1}$} &  \multicolumn{2}{|c|}{$E_\mu$ -- $[\hbar]^{\frac{1}{2}}[c]^{-\frac{1}{2}}$}  & $F$ -- \\  \hline
%%%%%%%
$H_{\!\ssc G\!H}(3)$	&	$t$ -- {$[\hbar]^{\frac{1}{2}}[c]^{-\frac{1}{2}}$}   & $x^i$ --	{$[\hbar]^{\frac{1}{2}}[c]^{\frac{1}{2}}$} 	& $\omega^{ij}$	 & $v^i$ -- $[c]$  & $p^i$ --  {$[\hbar]^{\frac{1}{2}}[c]^{-\frac{1}{2}}$} & $e$ -- {$[\hbar]^{\frac{1}{2}}[c]^{\frac{1}{2}}$}    &  $g$ -- $[\hbar]$\\ 
	& $T$ -- $[\hbar]^{-\frac{1}{2}} [c]^{-\frac{1}{2}}$  & $X_{\ssc i}$ -- $[\hbar]^{-\frac{1}{2}}[c]^{\frac{1}{2}}$    & $J_{ij}$   & $K_i$ --  $[c]^{-1}$   & $P_i$  -- $[\hbar]^{-\frac{1}{2}}[c]^{-\frac{1}{2}}$  & {$H$  -- $[\hbar]^{-\frac{1}{2}}[c]^{\frac{1}{2}}$} &  $G$ -- $[\hbar]^{-1}$  \\ 	
 $\times  [\hbar]$ & $T$ -- {$[\hbar]^{\frac{1}{2}}[c]^{-\frac{1}{2}}$}   & $X_{\ssc i}$ -- {$[\hbar]^{\frac{1}{2}}[c]^{\frac{1}{2}}$}     & $J_{ij}$ -- $[\hbar]$   & $K_i$ --  $[\hbar][c]^{-1}$   & $P_i$  -- $[\hbar]^{\frac{1}{2}}[c]^{-\frac{1}{2}}$  & {$H$  -- $[\hbar]^{\frac{1}{2}}[c]^{\frac{1}{2}}$} &  $G$ --   \\ 	  \hline
%%%%
$S_{\ssc G}(3)$	&	$t_c$ -- {$[\hbar]^{\frac{1}{2}}[c]^{-\frac{1}{2}}[k_p]$}	& $x^i_c$ --	{$[\hbar]^{\frac{1}{2}}[c]^{\frac{1}{2}}[k][k_p]$}    &  $\omega^{ij}$	& $v_c^i$ -- $[ck]$ & $p^i_c$ --  {$[\hbar]^{\frac{1}{2}}[c]^{-\frac{1}{2}}[k_x]$} 	 & $e_c$ -- {$[\hbar]^{\frac{1}{2}}[c]^{\frac{1}{2}}[k_x]$}  & \\  
		&  $T^c$  -- $[\hbar]^{-\frac{1}{2}}[c]^{-\frac{1}{2}}[k_x]^{-1}$ &  $X_i^{\!c}$  -- $[\hbar]^{-\frac{1}{2}}[c]^{\frac{1}{2}}[k_x]^{-1}$  & $J_{ij}$   & $K_i^c$ --  $ [c]^{-1}[k]^{-1}$ & $P^c_i$  -- $[\hbar]^{-\frac{1}{2}}[c]^{-\frac{1}{2}}[k]^{-1}[k_p]^{-1}$ &{$H^c$  -- $[\hbar]^{-\frac{1}{2}}[c]^{\frac{1}{2}} [k_p]^{-1} $} &  \\
 $\times  [\hbar][k][k_x] [k_p] $ &  $T^c$  -- $[\hbar]^{\frac{1}{2}}[c]^{-\frac{1}{2}}[k][k_p]$ &  $X_i^{\!c}$  -- $[\hbar]^{\frac{1}{2}}[c]^{\frac{1}{2}}[k][k_p]$  & $J_{ij}$  -- $[\hbar][k][k_x] [k_p] $ & $K_i^c$ --  $[\hbar] [c]^{-1}[k_x][k_p]$ & $P^c_i$  -- $[\hbar]^{\frac{1}{2}}[c]^{-\frac{1}{2}}[k_x]$ &{$H^c$  -- $[\hbar]^{\frac{1}{2}}[c]^{\frac{1}{2}}[k] [k_x] $} &  \\
   \hline
%%%
$S(1,3)$				&	\multicolumn{2}{|c|}{$\lambda^\mu_c$  -- {$[\hbar]^{\frac{1}{2}}[c]^{-\frac{1}{2}}[k_p]$}} & \multicolumn{2}{|c|}{$\omega^{\mu\nu}$}  	&\multicolumn{2}{|c|}{$p^\mu_c$ --  {$[\hbar]^{\frac{1}{2}}[c]^{-\frac{1}{2}}[k_x]$}}    &	$f$ -- $[\hbar][c]^{-1}$ 	\\
	&  \multicolumn{2}{|c|}{$X^c_\mu$ -- $[\hbar]^{-\frac{1}{2}}[c]^{\frac{1}{2}}[k_x]^{-1}$}  &  \multicolumn{2}{|c|}{$J_{\mu\nu}$}   &
\multicolumn{2}{|c|}{$E^c_\mu$ -- $[\hbar]^{-\frac{1}{2}}[c]^{\frac{1}{2}}[k_p]^{-1}$}  & $F$ --  $[\hbar]^{-1}[c]$  \\  
	$\times  [\hbar][c]^{-1}[k_x][k_p]$	&  \multicolumn{2}{|c|}{$X^c_\mu$ -- $[\hbar]^{\frac{1}{2}}[c]^{-\frac{1}{2}}[k_p]$}  &  \multicolumn{2}{|c|}{$J_{\mu\nu}$ -- $[\hbar][c]^{-1}[k_x][k_e]$}   &
\multicolumn{2}{|c|}{$E^c_\mu$ -- $[\hbar]^{\frac{1}{2}}[c]^{-\frac{1}{2}}[k_x]$}  & $F$ -- $[k_x][k_p]$ \\  \hline
%%%%%%
%$S_{\ssc G}(3)$	&	$t_c$ -- {$[\hbar]^{\frac{1}{2}}[c]^{-\frac{1}{2}}[k_p]$} & $x^i_c$ --$[\hbar]^{\frac{1}{2}}[c]^{\frac{1}{2}}[k_p]$	    &  $\omega^{ij}$	& $v^i$ -- $[c]$ & $p^i_c$ --  {$[\hbar]^{\frac{1}{2}}[c]^{-\frac{1}{2}}[k_x]$} 	 & $e_c$ -- {$[\hbar]^{\frac{1}{2}}[c]^{\frac{1}{2}}[k_x]$}  &  $f$ -- $[\hbar][c]^{-1}$\\ 
% $\times  [\hbar][k_x] [k_p]$  &  $X_{\!\ssc 0}$ -- $[\hbar]^{\frac{1}{2}}[c]^{\frac{1}{2}}[k_x] [k_p]$  & $X_{\ssc i}$ -- $[\hbar]^{\frac{1}{2}}[c]^{\frac{1}{2}}[k_x] [k_p]$    & $J_{ij}$   -- $[\hbar][k_x] [k_p]$ & $K_i$ --  $[\hbar][c]^{-1}[k_x] [k_p]$ & $P^c_i$  -- $[\hbar]^{\frac{1}{2}}[c]^{-\frac{1}{2}}[k_x]$ &{$H^c$  -- $[\hbar]^{\frac{1}{2}}[c]^{\frac{1}{2}}[k_x]$} & $F$ -- $[k_x] [k_p]$ \\
\hline\hline
\end{tabular}}
\end{center} 
%\end{table}
\end{sidewaystable}

%%%%%%%

\end{document}